\documentclass[12pt]{article}
\usepackage{amsmath,amssymb,amsfonts,epsfig}

\usepackage{bm}

\makeatletter\@addtoreset{equation}{section}\makeatother

\newcommand{\bep}{\left(\begin{array}{ccc}}
\newcommand{\bepr}{\left(\begin{array}{rrr}}
\newcommand{\eep}{\end{array}\right)}
\newcommand{\beB}{\left\{\begin{array}{ccc}}
\newcommand{\beBr}{\left\{\begin{array}{rrr}}
\newcommand{\eeB}{\end{array}\right\}}
\newcommand{\nn}{\nonumber}

\newcommand{\tr}{\text{tr}}
\newcommand{\trn}[2]{\text{tr}_{#1}\left( #2 \right)}
\newcommand{\bra}[1]{\left\langle #1 \right|}
\newcommand{\ket}[1]{\left| #1 \right\rangle}
\newcommand{\vev}[1]{\left\langle #1 \right\rangle}

\newcommand{\ylm}{Y_{lm}}
\newcommand{\tlm}{T_{lm}}
\newcommand{\tn}[1]{T_{l_{#1}m_{#1}}}
\newcommand{\tnd}[1]{T_{l'_{#1}\,m'_{#1}}}

\newcommand{\plm}{\phi_{lm}}

\newcommand{\pin}[1]{\phi^{\text{in}}_{l_{#1}m_{#1}}}
\newcommand{\pind}[1]{\phi^{\text{in}}_{l'_{#1}\,m'_{#1}}}

\newcommand{\phin}{{\phi^{\text{in}}}}

\newcommand{\pon}[1]{\phi^{\text{out}}_{2L\,m_{#1}}}
\newcommand{\pond}[1]{\phi^{\text{out}}_{2L\,m'_{#1}}}

\newcommand{\phout}{{\phi^{\text{out}}}}

\newcommand{\Lap}[1]{\left[L_i,\left[L_i,#1\right]\right]}

\newcommand{\Lo}{\Lambda_{\text{out}}}
\newcommand{\Li}{\Lambda_{\text{in}}}

\newcommand{\skinin}{S_N^{(\text{kin}.)\,\text{in}}}
\newcommand{\skinout}{S_N^{(\text{kin}.)\,\text{out}}}
\newcommand{\spot}{S_N^{(\text{pot}.)}}
\newcommand{\trN}[1]{\text{tr}_N \left( #1 \right)}
\newcommand{\kaN}[1]{\kappa_N^{#1}}

\allowdisplaybreaks

\input epsf

\begin{document}

\thispagestyle{empty}
\vspace*{-2em}
\begin{flushright}
%
\end{flushright}
\vspace{0.3cm}
\begin{center}
\Large {\bf
Existence of new nonlocal field theory \\
on noncommutative space and spiral flow \\
in renormalization group analysis of matrix models}

\vspace{0.7cm}

\normalsize
 \vspace{0.4cm}

Shoichi {\sc Kawamoto}$^{a,}$\footnote{e-mail address:\ \ 
{\tt kawamoto@cycu.edu.tw, kawamoto@yukawa.kyoto-u.ac.jp}}
and
Tsunehide {\sc Kuroki}$^{b,}$\footnote{
Since April 2015: {\it National Institute of Technology, Kagawa College, 551 Kohda, Takuma-cho, Mitoyo, 
Kagawa 769--1192, Japan.}\\
\hskip2em
e-mail address:\ \ 
{\tt kuroki@kmi.nagoya-u.ac.jp, kuroki@dg.kagawa-nct.ac.jp}}

\vspace{0.7cm}

$^a$ 
{\it
Department of Physics, Chung-Yuan Christian University,\\
Chung-Li 320, Taiwan, R.O.C.}\\

\vspace{0.4cm}

$^b$
{\it Kobayashi-Maskawa Institute for the Origin of Particles and the Universe, \\
Nagoya University, Nagoya 464-8602, Japan.}\\

\vspace{1cm}
{\bf Abstract}
\end{center}
In the previous study\cite{Kawamoto:2012ng,Kawamoto:2013dra},
we formulate a matrix model renormalization group based on the fuzzy spherical harmonics 
with which a notion of high/low energy can be attributed to matrix elements, 
and show that it exhibits locality and various similarity to
the usual Wilsonian renormalization group of quantum field theory.
In this work, we continue the renormalization group analysis of a matrix model 
with emphasis on nonlocal interactions where the fields on antipodal points are coupled. 
They are indeed generated in the renormalization group procedure 
and are tightly related to the noncommutative nature of the geometry. 
We aim at formulating renormalization group equations including such nonlocal 
interactions and finding existence of nontrivial field theory with antipodal interactions 
on the fuzzy sphere.
We find several nontrivial fixed points and calculate the scaling dimensions
associated with them. 
We also consider the noncommutative plane limit and then no consistent fixed point is found. 
This contrast between the fuzzy sphere limit and the noncommutative plane limit would be 
manifestation in our formalism of the claim given by Chu, Madore and Steinacker 
that the former does not have UV/IR mixing, while the latter does.

\newpage
\setcounter{page}{1}

\tableofcontents

\setcounter{footnote}{0}

\section{Introduction}
\label{sec:intro}

Matrix models have been considered as constructive definition
of quantum gravity.
Relatively simple one- and two-hermitian matrix models
have been solved exactly and provide nonperturbative
formulation of two-dimensional Euclidean quantum gravity coupled to
$c<1$ conformal matters \cite{Knizhnik:1988ak,Brezin:1990rb}.
Yang-Mills types of matrix models with fermionic symmetry
have been proposed as nonperturbative formulation of 
string/M theory \cite{Banks:1996vh}.
Each of them is given as a lower dimensional reduction of maximally supersymmetric
Yang-Mills theory and have not yet been solved exactly as in the case of
two-dimensional quantum gravity.

The large-$N$ renormalization group (RG) method proposed by
Brezin and Zinn-Justin \cite{Brezin:1992yc} is an analytic
approach to investigate critical behavior of such a matrix model
in the large-$N$ limit.
For simple one- or two-hermitian matrix models
where exact results are available,
it has been shown that large-$N$ RG analysis
captures critical behavior of the models, first qualitatively
 \cite{Brezin:1992yc}
and later even quantitatively with the help of Schwinger-Dyson
equations \cite{Higuchi:1993he}.

Though the original motivation of this large-$N$ RG method is to analyze
models that we cannot solve exactly, and to find a way to overcome the so-called
$c=1$ barrier, the actual application has not been
explored much so far.\footnote{%
One may see \cite{Narayan:2002gv,Kawamoto:2013laa} for some trials.}
In the past decades, the focus of the study on large-$N$
matrix models as quantum gravity has been shifted to reduced supersymmetric
Yang-Mills type matrix models as mentioned above.
There have been lots of study on these models 
(see e.g. \cite{Nishimura:2001sx}), 
including very impressive numerical studies (e.g. \cite{Kim:2011cr}), 
but it is of course very plausible if we can develop
an analytic approach to dissect such models.
One of the characteristics of these ``new'' matrix models is 
that the matrix elements
have direct physical interpretation; for example, they are often interpreted as
positions of D-branes and open strings connecting them.
On the other hand, in the case of matrix models for
Euclidean $D<2$ quantum gravity, the matrices are technical tools
to sum up random surfaces with spin degrees of freedom that give rise to
conformal matters.
Thus, the large-$N$ RG method proposed by Brezin and Zinn-Justin 
somehow inherits
this nature; namely, there are no criteria on which matrix elements are
to be integrated out.
On the other hand, in the modern interpretation, we may want to attribute
some physical meaning to matrix elements, and formulate large-$N$ RG based on 
this.
With this motivation, together with D.~Tomino, we have formulated a large-$N$
RG by using the fuzzy spherical harmonics \cite{Hoppe}. 
The fuzzy spherical harmonics are an analogue of the standard spherical harmonics,
and they span a basis on which general hermitian matrices can be expanded.
Thus, there appears a clear notion of high/low energy modes for expansion coefficients.
We can in this way formulate the large-$N$ RG, and have shown that they enjoy
several nice properties such as locality and derivative expansion of double trace terms.
We also find Gaussian and non-Gaussian fixed points,
and discuss their properties.
However, we have also faced a difficulty that originates in a noncommutative nature
of the geometry; it is embodied as emergence of new nonlocal interactions between 
fields on antipodal points on the fuzzy sphere. 
In the previous work, we carry out the RG analysis by dropping these antipodal interactions.
In this paper, we come back to this issue and present more complete treatment
of RG analysis.

In the following subsection, we start our discussion
with what we have done in the previous paper
and what we want to do in this paper.

\subsection{Large-$N$ renormalization group based on fuzzy spherical harmonics}
\label{sec:review_KKT}

We first review the basic formulation of the large-$N$ RG based on the fuzzy spherical harmonics.\footnote{%
Here only basics which are necessary to explain the problems
are presented. More detailed explanation is provided later.}
We then describe its intriguing aspects that have been observed in
the previous study \cite{Kawamoto:2012ng}.

We start with a matrix model that describes $\phi^4$
scalar field theory on a fuzzy $S^2$ of radius $\rho_N$,\footnote{%
In Appendix \ref{sec:matr-model-repr}, we summarize the basic properties
of the fuzzy sphere and also construction of the matrix model for
scalar field theory on it.}
\begin{align}
S_N=\frac{\rho_N^2}{N}\trn{N}{-\frac{1}{2\rho_N^2}[L_i,\phi]^2+\frac{m_N^2}{2}\phi^2+\frac{\kappa_N^{(0)}}{4}\phi^4},
\label{action}
\end{align}
where $N$ denotes the size of the hermitian matrix $\phi$, $L_i$ is the generator of the $SU(2)$ in the spin $L$ representation 
with $N=2L+1$, and $m_N^2$ and $\kappa_N^{(0)}$ are
the mass (squared) and the coupling constant respectively.\footnote{%
In \cite{Kawamoto:2012ng}, $\kappa_N^{(0)}$ was called $g_N$.
In this paper, the symbol $g_N$ is reserved for other combinations of the coupling constants.
In later sections, $\kappa_N^{(0)}$ stands for the coupling constant for the same operator $\tr_N \phi^4$.}
$\phi$ is expanded by use of $N\times N$ fuzzy spherical harmonics matrices,
\begin{align}
\phi=\sum_{l=0}^{2L}\sum_{m=-l}^l\plm \tlm.
\end{align}
The maximum of the angular momentum $2L$ is related to the size of the matrix
$N$ as $N=2L+1$.
The expansion coefficients $\plm$ are considered to be dynamical degrees of
freedom in field theory of $\phi$, and thus there exists a clear notion of ``high energy'' modes.
The large-$N$ RG transformation is defined by
integrating over $2L+1$ number of maximum angular momentum modes, $\phi_{2L m}$
with $-L \leq m \leq L$.
This can be carried out by usual perturbation theory, and it generates 
$\tr\,\phi^2$ and $\tr\,\phi^4$ terms (of the reduced size).
Upon a suitable change of the matrix basis, these changes can be absorbed into
the mass and the coupling constant.
This is an analogue of Wilsonian RG flow, and we can 
find Gaussian as well as non-Gaussian fixed points in the previous study \cite{Kawamoto:2012ng}. 
In fact, this procedure can also be regarded as the Wilsonian RG of a scalar field theory on the fuzzy sphere. 
We also examined linearized RG transformation around the fixed points and calculated the scaling
dimensions of operators at the fixed points as its eigenvalues.
Various large-$N$ limits with different noncommutativity respected are also
argued.

The perturbative calculation of the RG transformation also generates various terms that are not present 
in the starting action \eqref{action};
up to $\mathcal{O}\big( (\kappa_N^{(0)})^2 \big)$
 we considered, it includes
$\tr_N \phi^6$ term and also double trace terms.
It turns out that in the low energy regime $l\ll L$ these terms 
are either negligible or
can be rewritten as a series of single trace operators with derivatives
(commutators with $L_i$'s)
which are suppressed in $1/N$.
Namely, it has been observed that the corrections are controlled as in the case
of usual Wilsonian RG analysis.
However, it is found that there appears another type of correction terms
that does not present in the usual Wilsonian RG.
Let us consider the following simplest example,
\begin{align}
\tr_N  \vev{\phi \phout \phi \phout} \propto
\sum_{l=0}^{2L-1} \sum_{m=-l}^l
(-1)^l \phi_{lm} \tr_N \big( \phi T_{lm} \big)
\bigg[ 1+ \mathcal{O} \bigg(\frac{1}{L} \bigg) \bigg]
 \,,
\end{align}
where $\phout= \sum_m \phi_{2L m} T_{2L m}$ is the highest momentum mode
to be integrated out and the other two $\phi$ are low energy modes with $l<2L$.
The expectation value means that $\phout$ are contracted 
by use of the tree level propagator.
$1/L$ corrections can be expressed as derivatives on $\phi$,
but it is sufficient to look at the leading term for the current purpose.
If  $(-1)^l$ is absent, the leading term is simply $\tr_N \phi^2$,
but it cannot be organized in that way due to this phase factor.
Inspired by the parity property of the spherical harmonics,
\begin{align}
  Y_{lm}(\theta,\varphi) =& (-1)^l Y_{lm}(\pi-\theta, \varphi+\pi) \,,
\end{align}
 we may include the phase factor by the definition of a new matrix,
 \begin{align}
   \phi^A =& \sum_{l,m} (-1)^l \phi_{lm} T_{lm} \,,
 \end{align}
and call it an antipode matrix, or an 
antipode field on the fuzzy sphere,\footnote{%
It is possible to define the antipode basis $T^A_{lm} = (-1)^l T_{lm}$
and write $\phi^A = \sum \phi_{lm} T^A_{lm}$.} as the aforementioned property
implies that this field resides at the antipodal point on $S^2$.
Thus, the leading term is written as $\tr_N (\phi \phi^A)$, which can be seen
as the most nonlocal two-point interaction term on $S^2$. 
In fact, by using \eqref{Tortho} and \eqref{intandtr}
 in Appendix \ref{sec:matr-model-repr}, we find 
\begin{align}
\frac1N\trn{N}{\phi\phi^A}=\int\frac{d\Omega}{4\pi}\phi(\theta,\varphi)\phi(\pi-\theta,\varphi+\pi).
\label{eq:phiphiA}
\end{align}
Actually, integrating out the highest modes generates various types of new nonlocal interaction
terms with antipode fields. It is worth noticing that this nonlocality is different from 
what arises in the usual noncommutative field theory defined by the star product.\footnote[1]{%
Such nonlocality may reflect a stringy nature of the theory.
In a slightly different context, a string-like degree of freedom that connects
two opposite points on a squashed fuzzy sphere has been studied in \cite{Andronache:2015sxa}.} 
Indeed, according to \eqref{star} in Appendix \ref{sec:matr-model-repr}, 
a noncommutative product between functions $\phi_1(\theta,\varphi)*\phi_2(\theta,\varphi)$ 
corresponds to the matrix product $\phi_1\phi_2$, which is different from 
$\phi_1(\theta,\varphi)\phi_2(\pi-\theta,\varphi+\pi)$ corresponding to the matrix product 
$\phi_1\phi_2^A$. 
In the previous paper, we simply drop these new terms, and carry out RG analysis.

However, in the first place, according to the spirit of RG, we have to include interaction terms 
with the antipode fields from the beginning because they are generated in the RG. 
Furthermore, as well known in quantum field theory 
on noncommutative geometries (we call it noncommutative field theory or NCFT),
such an IR effect due to the UV loop integral is a characteristic feature of NCFT 
(UV/IR mixing \cite{Minwalla:1999px}), 
and it is then of great importance and interest to investigate them in further details.
If the terms with antipode fields are kept, the resultant effective action
is not a smooth function of the momentum $l$ even in the low energy regime,
due to the oscillating sign.
In \cite{Vaidya:2001bt}, Vaidya has argued that this is due to UV/IR mixing effect and concluded that the Wilsonian RG cannot be implemented in scalar field theory on the fuzzy sphere.
However, Chu, Madore, and Steinacker have criticized this conclusion by showing that
integrating out the all momenta, instead of just $l=2L$, leads to much milder behavior
and the two-point function is smooth at small values of the external momenta $l$ \cite{Chu:2001xi}.
They have also shown that 
a contribution to the two-point function of noncommutativity, which they call \textit{noncommutative anomaly}, remains finite even after taking a continuum $S^2$ limit.
Thus this reflects the effect from the underlying noncommutativity.
This noncommutative anomaly is also shown to lead the usual UV/IR mixing effect
if we take a noncommutative plane limit by blowing up a point on the fuzzy sphere.
In the context of the large-$N$ RG analysis, this observation suggests that we would have
a well-defined RG flow by performing RG transformation repeatedly.
If so, the large-$N$ RG will open up a possibility that noncommutative field theories
are formulated constructively through large-$N$ matrix models.\footnote{%
Since nonlocal interactions are introduced,
apart from the noncommutative anomaly, there may be further points to be checked,
such as unitarity or causality (in Lorentzian case, for example \cite{Addazi:2015dxa}),
for field theory to be well-defined.
We leave this point for future study.}
In this paper, we pursuit such possibilities and will consider the large-$N$ RG with
antipode fields included.
\\[1em]

This paper is organized as follows: in the following section 
we formulate the large-$N$ RG based on the fuzzy spherical harmonics with antipode fields.
By integrating out high-momentum modes perturbatively, 
the RG equations are derived.
In Section \ref{sec:fp}, we find various fixed points and also carry out linearized
study around them.
There, we encounter a peculiar feature of the large-$N$ RG with antipode fields;
namely, the RG equations  are quite different for
the cases with the starting size of $N$ being even or odd.
Irrespective of this, it turns out that the position of fixed points and the critical exponents 
for each fixed points agree in even and odd $N$ cases.
Thus, it suggests that out large-$N$ RG analysis correctly captures \textit{universal}
features of the theory.
Finally, in Section \ref{sec:discussions},
we provide conclusions and discussions. 
In Appendix \ref{sec:matr-model-repr}, we summarize the basics
of matrix model formulation for scalar field theory on a fuzzy sphere
and enumerate useful formulas. 
Some details of perturbative calculations are presented in Appendix \ref{app:calc_VEVs}.

\section{Large-$N$ renormalization group on fuzzy sphere with antipode fields}
\label{sec:LNRG_with_AP}

In this section, the formulation of the large-$N$ RG 
of the matrix model describing scalar field theory on the fuzzy sphere 
with antipode fields is presented.
In the first subsection,
we set up our action for scalar field theory on the fuzzy sphere
that also includes interaction vertices of antipode fields.
This is the starting point of our large-$N$ RG analysis.
We then carry out
a perturbative calculation that corresponds to a coarse-graining procedure
in the following subsection.
Finally, we explain
suitable rescaling of the traces and the radius of the sphere $\rho_N$,
with respect to various noncommutative parameters in the large-$N$ limit,
and we derive a set of RG equations for mass parameters and coupling constants.

\subsection{Action with antipode fields}
\label{sec:action}

In the previous work, the basic action \eqref{action} is used to formulate
the RG equations for $m_N^2$ and $\kappa_N^{(0)}$.
As discussed in Section \ref{sec:review_KKT}, we consider large-$N$ RG analysis
with antipode fields.
Thus, we take the following extended action as our starting point,
\begin{align}
  S_N=&
S_N^{\text{(kin.)}}+S_N^{\text{(pot.)}}, \nn \\
 S_N^{\text{(kin.)}}=&
\frac1N\trN{-\frac12[L_i,\phi]^2
+\frac{\rho_N^2m_N^2}{2}\phi^2
-\frac{\zeta_N}{2}[L_i,\phi][L_i,\phi^A]
+\frac{\rho_N^2\tilde{m}_N^2}{2}\phi\phi^A}
\,, \nn \\
S_N^{\text{(pot.)}}=
&\frac{\rho_N^2}{4N}\trN{\kaN{(0)}\phi^4+\kaN{(1)}\phi^3\phi^A+\kaN{(2\alpha)}\phi^2(\phi^{A})^2
 +\kaN{(2\beta)}(\phi\phi^A)^2} \,,
\label{action_AP}
\end{align}
where $m_N^2$ and $\tilde{m}_N^2$ are mass parameters,
$\zeta_N$ a parameter for the kinetic term with an antipode field,
and $\kappa_N^{(a)}$ coupling constants.
The index $a$ takes the values
$0,1, 2\alpha$, and $2\beta$ to distinguish the types of the vertices, and in particular, 
their numbers represent how many antipode fields are contained in the vertices.
As shown in Appendix \ref{sec:prop-antip-proj}, 
the antipode projection $\phi\mapsto\phi^A$ has a property that all the fields inside a trace
can be antipode projected by reversing the order of them
inside the trace. See \eqref{adj_prop_AP}.
Thus, without loss of generality, we can restrict the number of the antipode
fields in a trace to be no more than half of the number of the fields. 
Hence \eqref{action_AP} is the most general action with the rotational symmetry $SO(3)$ 
on the fuzzy sphere and $\bm Z_2$ symmetry $\phi_{lm}\rightarrow-\phi_{lm}$
up to the second derivative in quadratic terms and 
${\cal O}(\phi^4)$ non-derivative interactions. 
It should be obvious from this observation that
introducing $\phi^A$ does not mean introduction of new
degrees of freedom, since
the actual degrees of freedom are the mode coefficients
$\phi_{lm}$.
This extended action is considered to be
just an addition of momentum dependent interaction terms.
On the other hand, it is much simpler to use $\phi^A$ to express the action with such interactions 
since it enables us to write them in terms of the matrix product and trace.
Thus, in the action we introduce the antipode field $\phi^A$ as if it is an independent field, 
but we note that we do not have to integrate $\phi^A$ as an independent variable of $\phi$ 
in the large-$N$ RG.

\subsection{Coarse-graining: perturbative calculation}
\label{sec:coarse-graining}

In the idea of the large-$N$ RG, the modes with high $l$ will be integrated out
to produce an effective action for the modes with lower angular momenta.
We may start with a generic situation; namely, $\hat{n}$ number of the outermost
momentum shells, $l=2L, 2L-1,  \cdots, 2L-\hat{n}+1$, are integrated out.
Thus, we first
divide the space of the angular momentum $\Lambda$ into
the one for higher modes to be integrated (called \textit{out} modes),
and the others (called \textit{in} modes) as
\begin{align}
&\Lambda=\left\{(l,m)\,|\,0\leq l\leq 2L,\,-l\leq m\leq l\right\}
=\Lo^{(\hat{n})} \oplus \Li^{(\hat{n})}
 , \nn \\
&\Lo^{(\hat{n})} =
\left\{(l,m)\,|\,2L-\hat{n}+1 \leq l\leq2L,\,-l \leq m\leq l\right\}, \nn\\
&\Li^{(\hat{n})} =
\left\{(l,m)\,|\,0\leq l\leq2L-\hat{n},\,-l \leq m\leq l\right\},
\label{def:mom_shell_n-step}
\end{align}
and define correspondingly 
\begin{align}
\plm=
  \begin{cases}
\phi^\text{out}_{lm} & (l,m)\in\Lo^{(\hat{n})} \\
\pin{} & (l,m)\in\Li^{(\hat{n})}    
  \end{cases}
\label{def:in-out-fields}
\,.
\end{align}
A matrix only with in-modes, $\phin$, and similarly $\phout$,
are defined as
\begin{align}
\phin=\sum_{(l,m)\in\Li^{(\hat{n})}}\pin{}\tlm, \quad 
\phout=\sum_{(l,m)\in\Lo^{(\hat{n})}}\phi^\text{out}_{lm} \tlm \,.
\end{align}
The coarse-graining procedure is then formulated as
\begin{align}
S_{N-\hat{n}}(m_{N-\hat{n}}^2,\tilde{m}_{N-\hat{n}}^2, \kappa^{(a)}_{N-\hat{n}}) =-\log\int\prod_{(l,m)\in \Lambda_\text{out}^{(\hat{n})}}d\phi^\text{out}_{lm} \,e^{-S_N(m_N^2,\tilde{m}_N^2,\kappa^{(a)}_N)},
\label{RG}
\end{align}
where $S_{N-\hat{n}}(m_{N-\hat{n}}^2,\tilde{m}_{N-\hat{n}}^2, \kappa^{(a)}_{N-\hat{n}})$
is an action of $(N-\hat{n})\times (N-\hat{n})$ matrix.
As in the previous study \cite{Kawamoto:2012ng}, the large-$N$ RG formalism 
presented here respects the rotational symmetry $SO(3)$, and the resultant
action should also be organized with respect to $SO(3)$ irreducible representations.
So far, we construct a mapping from $N\times N$ matrix to
$(N-\hat{n})\times (N-\hat{n})$ matrix.
Instead of integrating out $l=2L, 2L-1 , \cdots 2L-\hat{n}+1$ modes in one
coarse-graining procedure, we may repeatedly perform the smallest $\hat{n}=1$
mapping $\hat{n}$ times to realize the same mapping.
As we discuss in Section \ref{sec:2-step-rg}, truncation or approximations make difference 
between them in general, but it can be negligible to lower order calculations we consider.
Thus, we restrict ourselves to considering $\hat{n}=1$ case for the time being,
and establish large-$N$ RG equations. 

We stress here that our RG \eqref{RG} is not just a mathematical problem of interest, 
but has firm physical ground. In fact, the mapping rule 
\eqref{mapping}--\eqref{Lapandadj} given in Appendix \ref{sec:matr-model-repr} 
tells us that the action \eqref{action_AP} we start from is completely equivalent 
to a field theory on a fuzzy sphere with radius $\rho_N^2$ 
\begin{align}
S=\int\frac{\rho_N^2d\Omega}{4\pi}\biggl(&-\frac{1}{2\rho_N^2}
\Bigl({\cal L}_i\phi(\theta,\varphi)\Bigr)^2
+\frac{m_N^2}{2}\phi(\theta,\varphi)^2 \nn \\
&-\frac{\zeta_N}{2\rho_N^2}
{\cal L}_i\phi(\theta,\varphi)\,{\cal L}_i\phi(\pi-\theta,\varphi+\pi)
+\frac{\tilde m_N^2}{2}\phi(\theta,\varphi)\phi(\pi-\theta,\varphi+\pi) \nn \\
&+\frac{\kappa_N^{(0)}}{4}\phi(\theta,\varphi)^4
  +\frac{\kappa_N^{(1)}}{4}\phi(\theta,\varphi)^3\phi(\pi-\theta,\varphi+\pi)
  +\frac{\kappa_N^{(2\alpha)}}{4}\phi(\theta,\varphi)^2\phi(\pi-\theta,\varphi+\pi)^2
   \nn \\
  &+\frac{\kappa_N^{(2\beta)}}{4}
\Bigl(\phi(\theta,\varphi)\phi(\pi-\theta,\varphi+\pi)\Bigr)^2
 \biggr), 
\label{ftaction_AP}
\end{align}
as a function of $\phi_{lm}$. Here all products are understood 
as a noncommutative product defined in \eqref{star}.\footnote
{As is well known, ${\cal O}(\phi^2)$ or ${\cal O}(\phi\phi^A)$ terms are not affected 
by the noncommutativity and hence the product can be replaced with 
the usual one as in \eqref{eq:phiphiA}.} 
Hence \eqref{RG} can also be regarded as applying the standard Wilsonian RG 
to the field theory \eqref{ftaction_AP} on the fuzzy sphere with antipodal interactions, 
which is cut off in such a way that the rotational symmetry is preserved. 
Thus our RG also reveals properties of such a field theory. Furthermore, as we discuss later, 
our RG is formulated to describe the large-$N$ limit with noncommutativity $\alpha$ 
of the fuzzy sphere fixed (see \eqref{eq:alpha} in Appendix \ref{sec:matr-model-repr}). 
Therefore we can fix $\alpha$ as small as we like to describe a field theory on the sphere 
regularized in a rotationally invariant way by small enough noncommutativity.  

To carry out the integration for out modes by perturbation theory, 
we first decompose the quadratic part of the action
as
\begin{align}
S_N^{(\text{kin}.)}
=&
\skinin+\skinout, \nn \\
\skinin =&
\frac1N\trN{-\frac12[L_i,\phin]^2
+\frac{\rho_N^2m_N^2}{2}\phin^2
-\frac{\zeta_N}{2}[L_i,\phin][L_i,\phin^A]
+\frac{\rho_N^2\tilde{m}_N^2}{2}\phin\phin^A}
\,, \nn \\
\skinout =&
\frac1N\trN{-\frac12[L_i,\phout]^2
+\frac{\rho_N^2m_N^2}{2}\phout^2
-\frac{\zeta_N}{2}[L_i,\phout][L_i,\phout^A]
+\frac{\rho_N^2\tilde{m}_N^2}{2}\phout\phout^A}
\nn\\=&
\sum_{m=-2L}^{2L}\frac12 \big[
N(N-1) \big( 1 + (-1)^{N-1} \zeta_N \big)
+\rho_N^2 \big(m_N^2+(-1)^{N-1} \tilde{m}_N^2 \big) \big]
\phi^{\text{out}\,*}_{2L\,m}\pon{} \,.
\end{align}
Note that there is no cross term between $\phin$ and $\phout$ in the quadratic part
of the action.
Here, we have used the fact that for the out modes the antipode projection is
simply multiplying the phase factor $(-1)^{N-1}$,
\begin{align}
  \phout^A =& \sum_{m=-2L}^{2L} (-1)^{2L} \phi_{2L m} T_{2L m}
=(-1)^{N-1} \phout \,.
\end{align}
This fact is also useful to organize the interaction vertices below.

We define 
\begin{align}
Z_0&=\int\prod_{m=-2L}^{2L}d\pon{}\,e^{-\skinout}, \quad 
\vev{\cal O}_0=\frac{1}{Z_0}\int\prod_{m=-2L}^{2L}d\pon{}
{\cal O}\,e^{-\skinout},
\end{align}
then our RG equation \eqref{RG} becomes 
\begin{align}
&S_{N-1}(m_{N-1}^2,\tilde{m}_{N-1}^2, \kappa^{(a)}_{N-1})=-\log Z_0+\skinin-\log\vev{e^{-\spot}}_0 \,.
\label{RGbyVEV}
\end{align}
Thus the calculation of $S_{N-1}(m_{N-1}^2,\kappa^{(a)}_{N-1})$
amounts to evaluating 
$\vev{e^{-\spot}}_0$. 
Now, in order to carry out perturbative calculation, we reorganize the interaction
part of the action $\spot$ according to the number of $\phin$, $\phout$ as well as
the number of antipode projections,
\begin{align}
S_N^{(\text{pot.})}=& \frac{\rho_N^2}{N}
\bigg[
\sum_{a=0,1,2\alpha,2\beta}
g_{0N}^{(a)} {\cal V}_0^{(a)}
+\sum_{b=0,1\alpha,1\beta} g_{1N}^{(b)} {\cal V}_1^{(b)}
+\sum_{c=0,1} \bigg( g_{2N}^{(c)P} {\cal V}_2^{(c)P}
+g_{2N}^{(c)NP} {\cal V}_2^{(c)NP} \bigg)
\nn\\& \hskip2em
+g_{3N}^{(0)} {\cal V}_3^{(0)}
+g_{4N}^{(0)} {\cal V}_4^{(0)}
\bigg] \,,
\end{align}
where
\begin{align}
 &{\cal V}_0^{(0)}=
\frac14\trn{N}{\phin^4} \,,
\quad
{\cal V}_0^{(1)} =
\frac14\trn{N}{\phin^3 \phin^A} \,, \nn \\
&\qquad \qquad \qquad \qquad \qquad 
{\cal V}_0^{(2\alpha)} =\frac14\trn{N}{\phin^2 (\phin^A)^2} \,,
\quad
{\cal V}_0^{(2\beta)} =
\frac14\trn{N}{(\phin \phin^A)^2} \,, \label{V0}
\\
 &{\cal V}_1^{(0)}=
\trn{N}{\phin^3 \phout} \,,
\quad
{\cal V}_1^{(1\alpha)} = \frac12\trn{N}{\phout \big(\phin^2 \phin^A
+ \phin^A\phin^2 \big)} \,, \nn \\
& \qquad \qquad \qquad \qquad \qquad \quad 
{\cal V}_1^{(1\beta)} = \trn{N}{\phout \phin \phin^A \phin} \,,
\label{V1}
\\
&{\cal V}_2^{(0)P} = \trn{N}{\phin^2 \phout^2} \,,
\quad
{\cal V}_2^{(0)NP} = \frac12\trn{N}{\left(\phin \phout\right)^2} \,,
\label{V20}
\\
&{\cal V}_2^{(1)P} = \frac12\trn{N}{\phout^2 \big(\phin\phin^A
+\phin^A\phin \big)} \,,
\quad
{\cal V}_2^{(1)NP} = \frac12\trn{N}{\phout \phin \phout \phin^A} \,,
\label{V21}
\\
 &{\cal V}_3^{(0)}=
\trn{N}{\phin \phout^3} \,, &&
\\
 &{\cal V}_4^{(0)}=
\frac14\trn{N}{\phout^4} \,, && \label{V4}
\end{align}
and the combinations of the original couplings are packed into $g_{iN}^{(a)}$
as
\begin{align}
 &g_{0N}^{(a)} = \kappa_N^{(a)} \,, \qquad
(a=0,1,2\alpha,2\beta) \\
 &g_{1N}^{(0)}=
\kappa_N^{(0)} + \frac14(-1)^{N-1} \kappa_N^{(1)} ,
\,\,
g_{1N}^{(1\alpha)}=
\frac12\kappa_N^{(1)} +  (-1)^{N-1}\kappa_N^{(2\alpha)}
 ,
\,\,
g_{1N}^{(1\beta)}=
\frac14\kappa_N^{(1)} +  (-1)^{N-1}\kappa_N^{(2\beta)}
\,,
\\
&g_{2N}^{(0)P}=
\kappa_N^{(0)} + \frac12 (-1)^{N-1} \kappa_N^{(1)}
+ \frac12 \kappa_N^{(2\alpha)} \,,
\qquad
g_{2N}^{(0)NP}=
\kappa_N^{(0)} +\frac12 (-1)^{N-1} \kappa_N^{(1)}
+ \kappa_N^{(2\beta)} \,,
\\
&g_{2N}^{(1)P} =
\frac12\kappa_N^{(1)} + \frac12(-1)^{N-1}\kappa_N^{(2\alpha)}
+ (-1)^{N-1}\kappa_N^{(2\beta)}
 \,,
\qquad
g_{2N}^{(1)NP} =
\frac12\kappa_N^{(1)} + (-1)^{N-1}\kappa_N^{(2\alpha)}
 \,,
\\
&g_{4N}^{(0)} =
g_{3N}^{(0)} =
\kappa_N^{(0)} 
+(-1)^{N-1} \kappa_N^{(1)} 
+\kappa_N^{(2\alpha)}  +\kappa_N^{(2\beta)} 
\,.
\label{g4}
\end{align}
The numbers in the lower indices for $\mathcal{V}_i^{(a)}$ and $g_{iN}^{(a)}$
stand for the number of out modes; namely the number of ``legs'' in the following
perturbation theory.
On the other hand, the numbers in the upper indices denote the number of
in fields with the antipode projection. The other labels, $\alpha$, $\beta$, and $P$ and $NP$, 
are for distinction of the types of vertices. The symmetry factors in \eqref{V0}--\eqref{V4} 
are assigned by paying attention to the fact that $\phi$ and $\phi^A$ are not independent 
as noted at the end of Section \ref{sec:action}. 
Notice that since the antipode projection on out fields simply provides an overall
alternating phase $(-1)^{N-1}$, only $\phin^A$ appears in the above vertices.
We have also utilized the trace property of antipode fields, \eqref{eq:ref_AP_trace},
to reduce the number of antipode fields.

The expectation value can be evaluated by using the propagator of the out modes 
\begin{align}
\vev{\pon{}\pond{}}_0=\delta_{m+m'}(-1)^mP_N,
\label{propagator}
\end{align}
where $P_N$ does not depend on $m$, $m'$ and has the form 
\begin{align}
P_N=\frac{1}{N(N-1)[1+(-1)^{N-1}\zeta_N] +\rho_N^2[m_N^2+(-1)^{N-1}\tilde{m}_N^2]}.
\label{propagator2}
\end{align}
By using this, we can perturbatively integrate out $\phout$.
It can be schematically summarized as
\begin{align}
-\log\vev{e^{-\spot}}_0=&
\frac{\rho_N^2}{N} \sum_{a=0,1,2\alpha,2\beta}
g_{0N}^{(a)} \mathcal{V}_0^{(a)}
+\frac{\rho_N^2}{N} 
\sum_{i=1}^4 \sum_{a} 
g_{iN}^{(a)} \vev{\mathcal{V}_i^{(a)}}_0
\nn\\&
-\frac{1}{2} \bigg( \frac{\rho_N^2}{N} \bigg)^2
\sum_{i,j=1}^4 \sum_{a,b}
g_{iN}^{(a)} g_{jN}^{(b)} \vev{\mathcal{V}_i^{(a)} \mathcal{V}_j^{(b)} }_c
+\mathcal{O}(\kappa_N^3) \,,
\end{align}
where $\vev{\cdots}_c$  means taking the connected part.
The summation over the indices $a$ and $b$ is understood as running over the possible
values including $P$ and $NP$ for $\mathcal{V}_2^{(a)}$ given in \eqref{V20} and \eqref{V21}.
$\mathcal{O}(\kappa_N^3)$ stands for the third or higher order corrections
in $\kappa_N^{(a)}$.
$\mathcal{V}_0^{(a)}$ does not contain $\phout$ and is then directly inherited to
$N-1$ theory.

The action \eqref{action_AP} has a $\mathbf{Z}_2$ symmetry $\phi \rightarrow -\phi$
(or more precisely, $\phi_{lm} \rightarrow -\phi_{lm}$), and so does $\skinout$.
Thus, expectation values that contain odd number of out modes vanish identically.
Furthermore, $\vev{\mathcal{V}_4^{(0)}}_0$ and $\vev{\mathcal{V}_4^{(0)}\mathcal{V}_4^{(0)}}_c$
does not include $\phin$ and they do not affect the renormalization of the parameters.
With these considerations, we have
\begin{align}
&  S_{N-1} \nn\\=&
S_N^\text{in}+\frac{\rho_N^2}{N} 
\sum_{a} g_{2N}^{(a)} \vev{\mathcal{V}_2^{(a)}}_0
\nn\\&
-\frac{1}{2} \bigg( \frac{\rho_N^2}{N} \bigg)^2
\bigg[
\sum_{a,b}
g_{1N}^{(a)} g_{1N}^{(b)} \vev{\mathcal{V}_1^{(a)} \mathcal{V}_1^{(b)} }_c
+2 \sum_{a}
g_{1N}^{(a)} g_{3N}^{(0)} \vev{\mathcal{V}_1^{(a)} \mathcal{V}_3^{(0)} }_c
+g_{3N}^{(0)} g_{3N}^{(0)} \vev{\mathcal{V}_3^{(0)} \mathcal{V}_3^{(0)} }_c
\nn\\& \hskip6em
+\sum_{a,b}
g_{2N}^{(a)} g_{2N}^{(b)} \vev{\mathcal{V}_2^{(a)} \mathcal{V}_2^{(b)} }_c
+2 \sum_{a}
g_{2N}^{(a)} g_{4N}^{(0)} \vev{\mathcal{V}_2^{(a)} \mathcal{V}_4^{(0)} }_c
\bigg]
\nn\\&
+\mathcal{O}(\kappa_N^3)
+(\text{$\phin$  independent terms})
-\ln Z_0  \,.
\label{eq:S_N-1_pert}
\end{align}
Here, $-\ln Z_0$ is also a $\phin$  independent term and we will no longer write
the last two terms explicitly.
$S_N^\text{in} =  \skinin
+\frac{\rho_N^2}{N} \sum_{a} g_{0N}^{(a)} \mathcal{V}_0^{(a)}$
 is the original action in \eqref{action_AP} with $\phi$ replaced with $\phin$.

We first argue that in the low energy regime of $\phin$ fields,
namely in which angular momenta $l$ associated with all of $\phin$ are much smaller
compared to $2L$, $l \ll L$, corrections involving $\mathcal{V}_1^{(a)}$ and
$\mathcal{V}_3^{(0)}$, \textit{i.e.} the second line in \eqref{eq:S_N-1_pert},
are negligible in the large-$N$ (therefore large-$L$) limit.
In order to evaluate these terms, it is sufficient to consider
\begin{align}
  \vev{\tr_N (\mathcal{O}_1 \phout) \tr_N(\mathcal{O}_2 \phout)}_c ,
\,\,
  \vev{\tr_N (\mathcal{O}_1 \phout) \tr_N(\phin \phout^3)}_c ,
\,\,
  \vev{\tr_N (\phin \phout^3) \tr_N(\phin \phout^3)}_c ,
\label{eq:pert_negligible}
\end{align}
where $\mathcal{O}_{i}$ ($i=1,2$) is a cubic order homogeneous polynomial
in $\phin$ and $\phin^A$.
For example, $\mathcal{V}_1^{(1\alpha)}$ corresponds to the choice
$\mathcal{O}_1=\frac{1}{2}\big( \phin^2 \phin^A + \phin^A \phin^2\big)$
 as in \eqref{V1}.
Since all three fields in $\mathcal{O}_i$ are in-modes, their angular momenta
$l_j$ ($j=1,2,3$) are all small compared to the cutoff, $l_j \ll L$.
The total angular momentum of $\mathcal{O}_i$ is, by the usual addition rule,
bounded by $l_1+l_2+l_3$ which is again much smaller than $L$.
The trace $\tr_N (\mathcal{O}_i \phout)$ is nonvanishing only when the momenta of
$\mathcal{O}_i$ and $\phout$ are equal, and 
this condition cannot be met.
Thus, this vertex does not contribute to the perturbative calculation
we consider now, and
the first two terms in \eqref{eq:pert_negligible} are indeed negligible.
The third term has already been considered in the previous study \cite{Kawamoto:2012ng}
as $\vev{V_3^2}_c$.
We simply quote the result as
\begin{align}
  \vev{\tr_N (\phin \phout^3) \tr_N(\phin \phout^3)}_c
=& (\text{polynomial in $L$}) \times  e^{3L \ln \frac{3}{4}} \,,
\label{eq:V3V3-result}
\end{align}
which is exponentially small for large-$L$.
Some more details are given in Appendix \ref{app:calc_VEVs}.

Therefore, we need to consider
\begin{align}
  &  S_{N-1}
\nn\\=&
S_N^{in}
+\frac{\rho_N^2}{N} 
\sum_{a=0,1} \biggl[
g_{2N}^{(a)P} \vev{\mathcal{V}_2^{(a)P}}_0
+g_{2N}^{(a)NP} \vev{\mathcal{V}_2^{(a)NP}}_0
\biggr]
\nn\\ &
-\frac{1}{2}\bigg( \frac{\rho_N^2}{N} \bigg)^2
\bigg[
g_{2N}^{(0)P} g_{2N}^{(0)P} \vev{\mathcal{V}_2^{(0)P} \mathcal{V}_2^{(0)P}}_c
+2 g_{2N}^{(0)P} g_{2N}^{(0)NP} \vev{\mathcal{V}_2^{(0)P} \mathcal{V}_2^{(0)NP}}_c
\nn\\& \hskip6em
+g_{2N}^{(0)NP} g_{2N}^{(0)NP} \vev{\mathcal{V}_2^{(0)NP} \mathcal{V}_2^{(0)NP}}_c
+ 2 g_{2N}^{(0)P} g_{2N}^{(1)P} \vev{\mathcal{V}_2^{(0)P} \mathcal{V}_2^{(1)P}}_c
\nn\\& \hskip6em
+2 g_{2N}^{(0)P} g_{2N}^{(1)NP} \vev{\mathcal{V}_2^{(0)P} \mathcal{V}_2^{(1)NP}}_c
+2 g_{2N}^{(0)NP} g_{2N}^{(1)NP} \vev{\mathcal{V}_2^{(0)NP} \mathcal{V}_2^{(1)NP}}_c
\nn\\ & \hskip6em
+2 g_{2N}^{(1)P} g_{2N}^{(0)NP} \vev{\mathcal{V}_2^{(1)P} \mathcal{V}_2^{(0)NP}}_c
+g_{2N}^{(1)P} g_{2N}^{(1)P} \vev{\mathcal{V}_2^{(1)P} \mathcal{V}_2^{(1)P}}_c
\nn\\ & \hskip6em
+2 g_{2N}^{(1)P} g_{2N}^{(1)NP} \vev{\mathcal{V}_2^{(1)P} \mathcal{V}_2^{(1)NP}}_c
+g_{2N}^{(1)NP} g_{2N}^{(1)NP} \vev{\mathcal{V}_2^{(1)NP} \mathcal{V}_2^{(1)NP}}_c
\nn\\ & \hskip6em
+2 g_{2N}^{(0)P} g_{4N}^{(0)} \vev{\mathcal{V}_2^{(0)P} \mathcal{V}_4^{(0)}}_c
+2 g_{2N}^{(0)NP} g_{4N}^{(0)} \vev{\mathcal{V}_2^{(0)NP} \mathcal{V}_4^{(0)}}_c
\nn\\ & \hskip6em
+2 g_{2N}^{(1)P} g_{4N}^{(0)} \vev{\mathcal{V}_2^{(1)P} \mathcal{V}_4^{(0)}}_c
+2 g_{2N}^{(1)NP} g_{4N}^{(0)} \vev{\mathcal{V}_2^{(1)NP} \mathcal{V}_4^{(0)}}_c
\bigg]
\nn\\&
+ \mathcal{O}(\kappa^3)
+ (\text{irrelevant or negligible})
\,.
\end{align}
In order to evaluate relevant expectation values, it is sufficient to consider
the following pieces,
\begin{align}
&  \vev{\tr_N \left({\cal O}_1 \; \phout \; {\cal O}_2 \; \phout \right)}_0
=
 (2N-1) \; P_{N} 
\tr_N \bigg[
{\cal O}_1 {\cal O}_2^A
-\frac{1}{N} {\cal O}_1 (-\Delta) {\cal O}_2^A
+{\cal O}(N^{-2})
\bigg] \,,
\label{eq:OoutOout}
\\
  &
\vev{\tr_N \big({\cal O}_1 \phout {\cal O}_2 \phout \big)
\tr_N \big(\phout^4 \big)}_c
\nn\\&=
N (2N-1)^2 P_{N}^3
\tr_N \bigg[
{\cal O}_1 {\cal O}_2^A
-\frac{1}{N} {\cal O}_1 (-\Delta) {\cal O}_2^A
+{\cal O}(N^{-2})
\bigg] 
\label{eq:OoutOout-out4}
\,,\\
&
\vev{ \tr_N \left({\cal O}_1 \phout {\cal O}_2 \phout \right)
\tr_N \left({\cal O}_3 \phout {\cal O}_4 \phout  \right)}_c
\nn\\&=
N(2N-1) P_{N}^2
\tr_N \bigg[
\mathcal{O}_1^A \mathcal{O}_2 \mathcal{O}_3^A \mathcal{O}_4
-\frac{1}{2N}
\bigg(
-\sum_i \Delta^{(i)} \big(\mathcal{O}_1^A \mathcal{O}_2 \mathcal{O}_3^A \mathcal{O}_4 \big)
+ \mathcal{O}_1^A \mathcal{O}_2 \Delta \big( \mathcal{O}_3^A \mathcal{O}_4 \big)
\bigg)
\nn\\& \hskip9em
+(\mathcal{O}_3 \leftrightarrow \mathcal{O}_4)
+\mathcal{O}(N^{-2})
\bigg]
\label{eq:OoutOout-OoutOout}
 \,,
\end{align}
where $\mathcal{O}_i$ ($i=1,2,3,4$) are polynomials of $\phin$ and $\phin^A$
or an identity $\mathbf{1}$,
and some exponentially small terms are neglected. 
$\Delta$ is defined as in \eqref{matrixLaplacian} and $\Delta^{(i)}$ acts only on 
$\mathcal{O}_i$ or $\mathcal{O}_i^{A}$. 
These formulas are derived in Appendix \ref{app:calc_VEVs} ($\hat{n}=1$ case).
By choosing suitable $\mathcal{O}_i$, we can represent the various types of vertices.
For example,
\begin{align}
  \mathcal{V}_2^{(1)P} =& \tr_N \left({\cal O}_1 \phout {\cal O}_2 \phout \right)
\quad \text{with} \quad
\mathcal{O}_1=\frac{1}{2}\big( \phin \phin^A + \phin^A \phin \big) ,
\quad 
\mathcal{O}_2 = \mathbf{1} \,.
\end{align}
Then we can apply the general formulas to evaluate the expectation values.
With these formulas, we find
\begin{align}
&  S_{N-1}
\nn\\
&=
-\frac{1}{2N }\trN{[L_i,\phin]^2
+\zeta_N [L_i,\phin][L_i,\phin^A]}
\nn\\&
+\frac{\rho_N^2}{2N}
\bigg[
m_N^2
+ B_1(N)
\bigg(g_{2N}^{(0)}  + \frac12 g_{2N}^{(1)NP} \bigg)
\bigg(1
-\rho_N^2 B_2(N) g_{4N}^{(0)}
\bigg)
\bigg]  \tr_N \big( \phin^2 \big)
\nn\\&
+\frac{\rho_N^2}{2N}
\bigg[
\tilde{m}_N^2
+ B_1(N)
\bigg( g_{2N}^{(1)P}  + \frac12 g_{2N}^{(0)NP}  \bigg)
\bigg( 1
-\rho_N^2 B_2(N)  g_{4N}^{(0)} \bigg)
\bigg]  \tr_N \big( \phin \phin^A \big)
\nn\\&
+\frac{\rho_N^2}{4N}
\bigg[
g_{0N}^{(0)}
-\rho_N^2 B_2(N)
\bigg( g_{2N}^{(0)P} + \frac12 g_{2N}^{(1)NP} \bigg)^2
\bigg] \tr_N \big( \phin^4 \big)
\nn\\&
+\frac{\rho_N^2}{4N}
\bigg[
g_{0N}^{(1)}
- 4\rho_N^2 B_2(N)
\bigg(
 g_{2N}^{(0)P}+ \frac12 g_{2N}^{(1)NP} \bigg)
\bigg(
g_{2N}^{(1)P}+ \frac12 g_{2N}^{(0)NP} \bigg)
\bigg] \tr_N \big( \phin^3 \phin^A \big)
\nn\\&
+\frac{\rho_N^2}{4N}
\bigg[
g_{0N}^{(2\alpha)}
- \rho_N^2 B_2(N)
\bigg( \bigg( g_{2N}^{(0)P}+\frac12 g_{2N}^{(1)NP} \bigg)^2
+\bigg(g_{2N}^{(1)P} + \frac12 g_{2N}^{(0)NP} \bigg)^2
-\frac14\big(g_{2N}^{(0)NP}\big)^2 
\bigg)
\bigg] \nn \\
&\hspace{2em}\times\tr_N \big( \phin^2 (\phin^A)^2 \big)
\nn\\&
+\frac{\rho_N^2}{4N}
\bigg[
g_{0N}^{(2\beta)}
- \rho_N^2 B_2(N)
\bigg( \bigg( g_{2N}^{(1)P}+\frac12 g_{2N}^{(0)NP}  \bigg)^2
+\frac14\big(  g_{2N}^{(0)NP} \big)^2
\bigg)
\bigg] \tr_N \big( \phin \phin^A \phin \phin^A \big)
\nn\\&
+ \mathcal{O}(\kappa^3)
+ (\text{irrelevant or negligible})
\,,
\label{eq:S_N-1_result1}
\end{align}
where subleading contributions of $1/N$ are dropped.
We have defined
\begin{align}
  B_1(N) =&
B_1(N; m_N^2, \tilde{m}_N^2) =
2(2N-1) P_N \,,
\qquad
  B_2(N) =
B_2(N; m_N^2, \tilde{m}_N^2) =
2(2N-1) P_N^2 \,.
\end{align}
As shown here, we sometimes omit the mass dependence from $B_1(N)$
and $B_2(N)$ to make expressions concise.
Note that the mass dependence comes through the propagator
factor $P_N$ given in \eqref{propagator2}.

The coefficients of each operator will be identified with
new mass and coupling parameters of the size $N-1$ theory.
However, the trace is still defined in the space of $N\times N$ matrices, 
and the length scale $\rho_N$ may also be renormalized in the spirit of
Wilsonian RG.
In the next subsection, we deal with them.

\subsection{Mapping the trace and rescaling}
\label{sec:rescaling}

The result of the perturbative calculation
\eqref{eq:S_N-1_result1} is yet to be identified with
a theory of $(N-1)\times (N-1)$ matrices.
The trace is still defined in $N \times N$ space,
and the matrix basis should be replaced with the one with a smaller size.
Furthermore, after integrating out the higher momentum modes, the range of the momenta 
is changed, and it needs to be scaled to the original range as in the standard RG of field theory.
This procedure involves the renormalization of the radius $\rho_N$.
Together with these procedures, in order to fix the overall scale,
we will normalize $\phin$ so that the kinetic term has the canonical normalization.
Then we can fix relations between the parameters
in the original theory and the renormalized theory.

We first consider the mapping of the trace in $N\times N$ matrix space
into $(N-1)\times (N-1)$ one.
This is essentially the same procedure we took in \cite{Kawamoto:2012ng}.
Let us write the basis for $N \times N$ as $T_{lm}^{(N)}$ ($0 \leq l \leq 2L$)
and $(N-1) \times (N-1)$ as $T_{lm}^{(N-1)}$ with
$l=0, \cdots, 2L-1$.
Since $\phi_{lm}^\text{in}$ does not have components $l=2L$, we can define
$\phi_{lm}^\text{in} = c \tilde{\phi}_{lm}$ with $0 \leq l \leq 2L-1$,
and also an $(N-1)\times (N-1)$ matrix as 
$\tilde{\phi} = \sum_l \tilde{\phi}_{lm} T_{lm}^{(N-1)}$.
Here, $c$ is a constant to be fixed.
By \eqref{Tortho} and \eqref{matrixLaplacian} in Appendix \ref{sec:matr-model-repr}, 
the kinetic term is mapped as
\begin{align}
  \frac{1}{N} \tr_N \bigg( -[L_i, \phin]^2 \bigg)
=& \frac{c^2}{N-1} \tr_{N-1} \bigg( -[\tilde{L}_i, \tilde{\phi}]^2 \bigg)
\,,
\end{align}
where $\tilde{L}_i$ is the $SU(2)$ generator of spin $L-\frac{1}{2}$ representation.
We require that the kinetic term stays canonical through the RG procedure,
and then $c$ is fixed to be 1.
The trace of the quadratic term also transforms simply as
\begin{align}
  \frac{1}{N} \tr_{N} \big( \phin^2 \big) = \frac{1}{N-1} \tr_{N-1} \big( \tilde{\phi}^2 \big)
\,.
\end{align}
It is easy to see that the same normalization applies if the trace involves
antipode ones.

On the other hand, the quartic vertices turns out to have a nontrivial
factor.
A trace with four matrices is written in terms of $3j$ and $6j$ symbols
by \eqref{4T1}.
A nontrivial $N$ dependence comes from $L=(N-1)/2$ in two $6j$ symbols,
and we apply the following recursion relation from \cite{Var}
\begin{align}
&
  \begin{Bmatrix}
    a & b & l \\
    L & L & L
  \end{Bmatrix}
= \frac{1}{\sqrt{(2L+a+1)(2L-a)(2L+b+1)(2L-b)}}
\nn\\&   \times
\bigg[
-2L\sqrt{(2L+l+1)(2L-l)}
  \begin{Bmatrix}
    a & b & l \\
    L-\frac{1}{2} & L-\frac{1}{2} & L-\frac{1}{2}
  \end{Bmatrix}
+\sqrt{a(a+1)b(b+1)}
  \begin{Bmatrix}
    a & b & l \\
    L & L & L-1
  \end{Bmatrix}
\bigg]
\,.
\label{eq:6j_shift2}
\end{align}
When $a+b+l=$even, by use of \eqref{eq:CG_shift} and \eqref{eq:6j_shift},
we find
\begin{align}
  \begin{Bmatrix}
    a & b & l \\
    L & L & L
  \end{Bmatrix}
=&
\bigg(-1+\frac{1}{2N}+\frac{1}{8N^2}
+\mathcal{O}(N^{-3})  \bigg)
  \begin{Bmatrix}
    a & b & l \\
    L-\frac{1}{2} & L-\frac{1}{2} & L-\frac{1}{2}
  \end{Bmatrix}
\label{eq:6j_shift3}
\,.
\end{align}
Here in quartic vertices $a$ and $b$ are the angular momenta of in modes, 
while $l$ is summed over as in \eqref{4T1}. 
When $a+b+l$ is not even, we may roughly take the second term in \eqref{eq:6j_shift2}
to be subleading.
This leads to a similar relation to \eqref{eq:6j_shift3}, but $\mathcal{O}(N^{-2})$ 
term depends on $a$, $b$, and $l$.
Thus it is natural to expect that the coefficient of $\mathcal{O}(N^{-2})$ term
is still $1/8$, independent of $a$, $b$, and $l$,
for this case due to the continuity, but we do not have
a concrete expression at this moment. 
However, at least in the low energy regime $a,b\ll L$, it is no doubt that the second term 
in \eqref{eq:6j_shift2} is of ${\cal O}(1/N^2)$ compared to the first term. 
In summary, we can write down a mapping formula for quartic vertices as
\begin{align}
\frac{1}{N}  \tr_N \big( T^{(N)}_{l_1 m_1} T^{(N)}_{l_2 m_2} T^{(N)}_{l_3 m_3} T^{(N)}_{l_4 m_4} \big)
= \bigg(1 +\mathcal{O}(N^{-2})  \bigg)
\frac{1}{N-1}
\tr_{N-1} \big( T^{(N-1)}_{l_1 m_1} T^{(N-1)}_{l_2 m_2} T^{(N-1)}_{l_3 m_3} T^{(N-1)}_{l_4 m_4} \big)
\,.
\end{align}

Next, we consider the rescaling of $\rho_N^2$.
In the usual Wilsonian RG of quantum field theory,
we integrate out the momentum $p$ for the interval
$\Lambda/b \leq p \leq \Lambda$ where $\Lambda$ is a cutoff
and $b>1$ is a number associated with RG transformation.
After integration, we perform a scale transformation $p \rightarrow bp$ to
get back to the original momentum space.
It thus involves scale transformations of all the dimensionful quantities;
especially fields get rescaled, and from this change we can read 
the scaling dimensions of the fields.
We define a fuzzy sphere counterpart of $b$, called $b_N$, as
\begin{eqnarray}
b_N^2\equiv \frac{\rho_N^2}{\rho_{N-1}^2}. 
\label{scale}
\end{eqnarray} 
So the question is how to define a scale for $N-1$ theory.
Since the fuzzy sphere preserves the rotation $SO(3)$ symmetry, 
the most natural invariant is the total angular momentum $L_i^2$, and
the momentum squared is given by dividing it by the radius squared. 
Thus we require 
\begin{align}
\frac{2L(2L+1)}{\rho_N^2}=\frac{2L(2L-1)}{\rho_{N-1}^2} \,,
\label{fsscale}
\end{align}
which gives $b_N^2=\frac{N}{N-2}$.
It should be noted that the relation between the fundamental scale $\alpha$ and
the fuzzy sphere radius $\rho_N$
\begin{align}
\rho^2_N = \frac{\alpha^2(N^2-1)}{4}
\label{eq:rhoandalpha}
\end{align} 
(see Appendix \ref{sec:matr-model-repr}), is preserved
by the same $\alpha$, up to $\mathcal{O}(N^{-2})$ corrections.
Namely, this limit is a large-$N$ limit with the characteristic scale of the fuzzy sphere
$\alpha$ fixed which describes noncommutativity as in \eqref{eq:alpha}.
In this sense, we call this large-$N$ limit the ``fuzzy sphere limit''. 
Thus as stressed in \eqref{ftaction_AP}, our RG provides nonperturbative 
information of the field theory with antipodal interactions on the fuzzy sphere 
with fixed noncommutativity $\alpha$.

As discussed in \cite{Kawamoto:2012ng}, we can think of another 
large-$N$ limit, which is related to
noncommutative field theory (NCFT) \cite{Minwalla:1999px, Szabo:2001kg} 
on the flat two-dimensional plane (see e.g. \cite{Chu:2001xi, Iso:2001mg}).
Thus, we call it the NCFT limit.
In this case, the large-$N$ limit is taken with the noncommutativity $\theta$ in NCFT fixed as 
\cite{Chu:2001xi} 
\begin{align}
N\rightarrow\infty \quad \text{with} \quad \theta=\frac{2\rho_N^2}{N}:~\text{fixed} \,,
\label{NCFTlimit}
\end{align}
which leads to $b_N^2=\frac{N}{N-1}$.
This limit is to zoom up a tiny part of the sphere (say, the north pole),
which will be approximated by a plane with a noncommutativity
$[\hat{y}_1, \hat{y}_2] = i \theta$. 
In this case the RG is expected to describe a field theory on this plane (Moyal plane, 
or noncommutative plane).

\subsection{Renormalization group equations}
\label{subsec:RGE}

After finishing the rescaling and the mapping of the traces,
\eqref{eq:S_N-1_result1} should be identified with an action of $(N-1)\times (N-1)$
size, up to negligible terms,
\begin{align}
  S_{N-1} =&
\frac{1}{N-1}\tr_{N-1} \bigg[ -\frac12[\tilde{L}_i,\tilde{\phi}]^2
+\frac{\rho_{N-1}^2m_{N-1} ^2}{2}\tilde{\phi}^2
-\frac{\zeta_{N-1}}{2}[\tilde{L}_i,\tilde{\phi}][\tilde{L}_i,\tilde{\phi}^A]
+\frac{\rho_{N-1}^2\tilde{m}_{N-1}^2}{2}\tilde{\phi}\tilde{\phi}^A
\nn\\& \hskip3em
+ \frac{\rho_{N-1}^2}{4}
\bigg( \kappa_{N-1}^{(0)}\tilde{\phi}^4
+\kappa_{N-1}^{(1)}\tilde{\phi}^3 \tilde{\phi}^A
+\kappa_{N-1}^{(2\alpha)}\tilde{\phi}^2 (\tilde{\phi}^{A})^2
+\kappa_{N-1}^{(2\beta)} \big(\tilde{\phi} \tilde{\phi}^{A} \big)^2
\bigg) \bigg]
 \,.
\end{align}
The identification of the parameters leads to the following RG equations (RGEs),
\begin{align}
\zeta_{N-1} =& \zeta_N \,,
\\
   m_{N-1}^2 =&
b_N^2 m_N^2
+b_N^2 B_1(N) \mathcal{X}^1_N(\kappa_N)
\big( 1 - \rho_N^2 B_2(N) g_{4N}^{(0)}
\big)
+\mathcal{O}(\kappa^3)
\label{eq:m_1-step}
\,,\\
   \tilde{m}_{N-1}^2 =&
b_N^2  \tilde{m}_N^2
+b_N^2 {B}_1(N) \mathcal{X}^2_N(\kappa_N)
\big( 1 - \rho_N^2 B_2(N) g_{4N}^{(0)}
\big)
+\mathcal{O}(\kappa^3)
\label{eq:mt_1-step}
\,,\\
\kappa_{N-1}^{(a)} =&
b_N^2 \kappa_N^{(a)}
-b_N^2\rho_N^2
{B}_2(N) \mathcal{Y}^{(a)}_N (\kappa_N)
+\mathcal{O}(\kappa^3)
\label{eq:kappa_a_1-step}
\,,
\end{align}
with $a,b=0,1,2\alpha$, and $2\beta$.
Here we have introduced the functions of the four-vector $\kappa_N=\big( \kappa_N^{(0)},\kappa_N^{(1)},\kappa_N^{(2\alpha)},\kappa_N^{(2\beta)} \big) $ as 
\begin{align}
\mathcal{X}^{(1)}_N(\kappa_N)  =&g_{2N}^{(0)P}+\frac12 g_{2N}^{(1)NP}
=\kappa_N^{(0)} + \frac{1+2(-1)^{N-1}}{4} \kappa_N^{(1)}
+\frac{1+(-1)^{N-1}}{2} \kappa_N^{(2\alpha)}
\,,
\label{X1} \\
\mathcal{X}^{(2)}_N(\kappa_N) =&g_{2N}^{(1)P}+\frac12 g_{2N}^{(0)NP} \nn \\
=& \frac{1}{2}\bigg(
\kappa_N^{(0)} + \frac{2+(-1)^{N-1}}{2} \kappa_N^{(1)}
+(-1)^{N-1} \kappa_N^{(2\alpha)} 
+\big(1+2(-1)^{N-1} \big) \kappa_N^{(2\beta)}
\bigg)
\,,\\
\mathcal{X}^{(3)}_N(\kappa_N) =&\frac12 g_{2N}^{(0)NP}
=\frac{1}{2} \bigg( \kappa_N^{(0)} + \frac{(-1)^{N-1}}{2}\kappa_N^{(1)} 
+ \kappa_N^{(2\beta)} \bigg)
\,,\\
\mathcal{Y}^{(0)}_N (\kappa_N) =& 
\big( \mathcal{X}^{(1)}_N(\kappa_N) \big)^2 \,,
\\
\mathcal{Y}^{(1)}_N (\kappa_N) =&
4 \mathcal{X}^{(1)}_N(\kappa_N)  \mathcal{X}^{(2)}_N(\kappa_N)  \,,
\\
\mathcal{Y}^{(2\alpha)}_N (\kappa_N) =& 
\big( \mathcal{X}^{(1)}_N(\kappa_N) \big)^2
+\big( \mathcal{X}^{(2)}_N(\kappa_N) \big)^2
-\big( \mathcal{X}^{(3)}_N(\kappa_N) \big)^2
 \,,
\\
\mathcal{Y}^{(2\beta)}_N (\kappa_N) =& 
\big( \mathcal{X}^{(2)}_N(\kappa_N) \big)^2
+\big( \mathcal{X}^{(3)}_N(\kappa_N) \big)^2
 \,.
 \label{Y2b}
\end{align}
Notice that the subscript $N$ for $\mathcal{X}_N^{(i)}$ and $\mathcal{Y}_N^{(a)}$
refers only to the alternating coefficients in their definitions.

We note that the parameter for $\phi \phi^A$ kinetic term $\zeta_N$ does not
receive any correction.
Thus, in the RG procedure, we can take $\zeta_N=\zeta_{N-1}=\cdots=\zeta$
and $\zeta$ can be arbitrary.
In the next section, we start the fixed point analysis,
and we will set $\zeta=0$ for convenience.

\section{Fixed point analysis}
\label{sec:fp}

Since the large-$N$ limit corresponds to performing our RG infinitely many times, 
it would be described by fixed points of the RG transformation. 
If they exist, for each fixed point we can also deduce the scaling dimensions of operators 
in the large-$N$ limit from linearized RG transformation around the fixed point 
because the scaling dimension is response to the scale transformation. 
In this section, we look for fixed points of the set of the RGEs. 
If they exist, it would be strong evidence that a theory as in \eqref{ftaction_AP} 
with the parameters given by them exists consistently and nonperturbatively. 
Notice that for this reason the existence of a fixed point is striking in itself 
because we now allow quite nonlocal antipodal interactions. 
We then consider linearized analysis around the fixed points
to determine the scaling dimensions.

\subsection{Fixed points for 1-step RGEs}
\label{sec:fixed-points-1}

We first consider fixed points for the set of equations
\eqref{eq:m_1-step}--\eqref{eq:kappa_a_1-step}.
They are RGEs for the RG transformation from $N\times N$ theory to
$(N-1)\times (N-1)$ theory, which corresponds to the case with $\hat{n}$,
the number of momentum shells to be integrated out, being 1.
Therefore, we call them 1-step RGEs.

Fixed points are obtained by setting
\begin{align}
  m^2_{N-1}=m_N^2 = m_*^2 \,,
\qquad
  \tilde{m}^2_{N-1}=\tilde{m}_N^2 = \tilde{m}_*^2 \,,
\qquad
\kappa_{N-1}^{(a)} = \kappa_N^{(a)}=\kappa_*^{(a)} \quad
(a=0,1,2\alpha,2\beta) \,,
\end{align}
and solving the relations for $m_*^2$, $\tilde{m}_*^2$, and $\kappa_*^{(a)}$.
As noted in the previous section, the parameter $\zeta_N$ does not get renormalized
and then we can consistently set $\zeta_N=\zeta_{N-1}= \cdots =\zeta$.
We restrict ourselves to fixed points with $\zeta=0$.

Note that the mass parameters in $P_N$, $B_1(N)$, and $B_2(N)$ are all
set to be fixed point values.
We thus introduce the following notation,
\begin{align}
  P_N^* =& \frac{1}{N(N-1) + \rho_N^2 (m_*^2 + (-1)^{N-1}\tilde{m}_*^2)} \,,
\nn\\
B_1^*(N)=& B_1(N;m_*^2, \tilde{m}_*^2)= 2(2N-1) P_N^* \,,
\nn\\
B_2^*(N)=& B_2(N;m_*^2, \tilde{m}_*^2)= 2(2N-1) {P_N^*}^2
\,.
\label{eq:P*B1*B2*}
\end{align}
To simplify the analysis, we use the following rescaled variables
\begin{align}
  m_*^2 =& \frac{x_*^{(1)}}{\rho_N^2 P_N^*} \,,
\quad
  \tilde{m}_*^2 = \frac{x_*^{(2)}}{\rho_N^2 P_N^*} \,,
\quad
 \kappa_*^{(a)} = \frac{b_N^2-1}{b_N^2 \rho_N^2 B_2^*(N)} y_*^{(a)} \,.
\label{def:rescaled_1step}
\end{align}
By use of them, the fixed point equations are written as
\begin{align}
  x_*^{(i)} =& - \mathcal{X}^{(i)}_N(y_*) \bigg[
1- \frac{b_N^2-1}{b_N^2} g_{4*}^{(0)}
\bigg] \qquad (i=1,2)
\label{eq:1-step_FP_x}
 \,,\\
0=& y_*^{(a)} - \mathcal{Y}_N^{(a)}(y_*) \qquad
(a=0,1,2\alpha,2\beta) \,,
\label{eq:1-step_FP_y}
\end{align}
where $\mathcal{X}^{(i)}_N(y_*)$ is given in \eqref{X1} with $\kappa^{(a)}_N$ replaced 
with $y_*^{(a)}$ for each $a$ and the same for $\mathcal{Y}_N^{(a)}(y_*)$,
and $g_{4*}^{(0)}= y_*^{(0)} + (-1)^{N-1}y_*^{(1)} +y_*^{(2\alpha)}+y_*^{(2\beta)}$ 
as in \eqref{g4}.
Since the definition of $P_N^*$ involves $m_*^2$ and $\tilde{m}_*^2$,
the rescaling condition restricts the form of $P_N^*$ as
\begin{align}
  P_N^* =& \frac{1-x_*^{(1)}-(-1)^{N-1}x_*^{(2)}}{N(N-1)} \,.
\label{eq:P_star_1step}
\end{align}
Since \eqref{eq:1-step_FP_y} depends only on $y_*^{(a)}$, not $x_*^{(i)}$,
one may find solutions to them.
Then, by \eqref{eq:1-step_FP_x}, $x_*^{(i)}$ are uniquely determined for a given
set of $y_*^{(a)}$.
Together with \eqref{eq:P_star_1step}, the relations \eqref{def:rescaled_1step}
determines the fixed points in terms of the original variables.
Thus, the question boils down to finding solutions to \eqref{eq:1-step_FP_y}.
It should also be noticed that the second term in the square bracket of \eqref{eq:1-step_FP_x} is $1/N$ suppressed compared to the first term, 1,
and then we may neglect that term to discuss leading order fixed points as long as $g_{4*}^{(0)}$ 
is of ${\cal O}(1)$ at most.

From the definitions of $\mathcal{Y}_N^{(a)}$, one can see that the four equations
\eqref{eq:1-step_FP_y} have purely numeric coefficients
which depends on $N$ only through an alternating sign factor $(-1)^{N-1}$.
Thus, solutions can be searched numerically, for $N$ being even or odd separately, and
all solutions will be of order 1.

For even $N$, we find the following fixed points (or lines),
\begin{align}
  \big( y_*^{(0)}, y_*^{(1)}, y_*^{(2\alpha)}, y_*^{(2\beta)} \big) =&
(0,0,0,0) \,, \quad
(0,0,-4,4) \,, \nn\\
&
\bigg(t, 4(t\pm \sqrt{t}),
2(t \pm \sqrt{t}) , 2 \pm 2\sqrt{t}+t \bigg)
\,,\nn\\ &
\bigg(
\frac{1}{4}\big( 2+t \pm \sqrt{t+1} \big) ,
t ,
\frac{t}{2} ,
1-\frac{t}{4} \frac{1\mp \sqrt{t+1}}{1\pm \sqrt{t+1}}
\bigg)
\,,\nn\\ &
\bigg(
\frac{1}{2}\big( 1+t \mp \sqrt{2t+1} \big),
2t , t ,
\frac{1}{2} \big(3+t \pm \sqrt{2t+1} \big)
\bigg)
\,,\nn\\ &
\bigg(
t \pm 2\sqrt{t-1} ,
-4+4t \pm 4\sqrt{t-1}, 
-2+2t \pm 2\sqrt{t-1},
t
\bigg) 
\,,
\end{align}
where the double sign corresponds in each solution and $t$ is a parameter.
Formally, any $t \in \mathbf{R}$ solves the equations,
but may be restricted to the range in which the values $y_*^{(a)}$ stay
real.
The first one is the Gaussian fixed point, which obviously leads
to $m_*^2=\tilde{m}_*^2=\kappa_*^{(a)}=0$.
It is curious that the even $N$ case allows one-parameter families of solutions,
and then there are infinite number of fixed points irrespective of the fact that 
we have four equations of four variables.
Note that the first two solutions are not included in the one-parameter
families.

On the other hand, for odd $N$, there are four fixed points found,
\begin{align}
    \big( y_*^{(0)}, y_*^{(1)}, y_*^{(2\alpha)}, y_*^{(2\beta)} \big) =&
(0,0,0,0) \,, \quad
\big( 16, -64, 28, 20 \big) \,,
\nn\\ &
\bigg(
\frac{136}{9} \pm \frac{32\sqrt{2}}{3} ,
-64 \mp \frac{136\sqrt{2}}{3} ,
\frac{272}{9} \pm \frac{64\sqrt{2}}{3} ,
\frac{170}{9} \pm \frac{40\sqrt{2}}{3}
\bigg)  \,.
\end{align}
One can check that only the common fixed point for $N$ even and odd cases
is Gaussian one.
So far, no parametric solution has been found.\footnote{%
As discussed in the following subsections, the comparison between
even and odd $N$ cases should be done in terms of the 
original parameters, $m_*^2$ and so on.
However, the structures of the set of fixed points are so different,
and we do not expect common fixed points that make sense physically.
}

These observations suggest that the RG flow is stable only around the Gaussian fixed point
since we are considering the RG transformation from $N\times N$ theory to $(N-1)\times (N-1)$
one.
The non-Gaussian fixed points (or lines) does not make sense since they are not really
``fixed'' along the RG transformations.
The structure of fixed points are quite different, 
as $N$ even ones include critical lines, but $N$ odd ones not.
This is in contrast to the case of $N \rightarrow (N-2)$ flow we analyze later,
where $N$ even and odd ones have the same number of isolated fixed points.

Now we consider a linearized analysis around the Gaussian fixed point.
Near Gaussian fixed point (namely, $m_N^2 ,\tilde{m}_N^2, \kappa_N^{(a)} \ll 1$
and the same for the parameters of $N-1$), at the leading order in the large-$N$ limit
we have
\begin{align}
  &m_{N-1}^2 =b_N^2 \bigg( m_N^2 + \frac{1}{N} \sum_a 
\delta^{(a)} \kappa_N^{(a)} \bigg) ,\quad
  \tilde{m}_{N-1}^2 = b_N^2 \bigg( \tilde{m}_N^2 + \frac{1}{N} \sum_a 
\tilde\delta^{(a)} \kappa_N^{(a)}\bigg)  \,,\\
&\kappa_{N-1}^{(a)} = b_N^2 \kappa_N^{(a)}.
\label{kappaRGGaussian}
\end{align}
Here $a=0,1,2\alpha,2\beta$ and
\begin{align}
&  \delta^{(0)}= 4 , \quad
  \delta^{(1)}= 1+2(-1)^{N-1} , \quad
  \delta^{(2\alpha)}= 2+2(-1)^{N-1} , \quad
  \delta^{(2\beta)}= 0 , \nn\\
&  \tilde\delta^{(0)}= 2 , \quad
  \tilde\delta^{(1)}= 2+(-1)^{N-1} , \quad
  \tilde\delta^{(2\alpha)}= 2(-1)^{N-1} , \quad
  \tilde\delta^{(2\beta)}= 2+4 (-1)^{N-1} \,.
\end{align}
The behavior of the coupling constants $\kappa_N^{(a)}$ is trivial.
We first look at critical lines of the mass parameters, $\bar{m}(\kappa_N)$
and $\bar{\tilde{m}}(\kappa_N)$ which are defined by the following difference equations,
\begin{align}
    m_{N-1}^2 - \bar{m}(\kappa_{N-1}^{(a)}) =&
b_N^2 \big( m_{N}^2 - \bar{m}(\kappa_{N}^{(a)}) \big) ,
\qquad
  \tilde{m}_{N-1}^2 - \bar{\tilde{m}}(\kappa_{N-1}^{(a)}) =
b_N^2 \big( \tilde{m}_{N}^2 - \bar{\tilde{m}}(\kappa_{N}^{(a)}) \big)
\,.
\label{massRGGaussian}
\end{align}
Eqs.~\eqref{kappaRGGaussian} and \eqref{massRGGaussian} manifest that all variables 
have eigenvalue $b_N^2$ of the RG transformation around the Gaussian fixed point. 
Since in the RG we have made the scale transformation with $p\rightarrow b_Np$ as discussed 
in \eqref{scale}, this implies they all have the scaling dimension 2. 
Under the assumption that $\bar{m}$ and $\bar{\tilde{m}}$
are linear functions of $\kappa_N^{(a)}$, these equations can be solved in the large-$N$
limit for even and odd $N$ cases separately,
\begin{align}
\text{Even $N$:}\quad &
  \bar{m}= \ln N \big( -4 \kappa_N^{(0)} + \kappa_N^{(1)} \big) \,,
\nn\\ &
 \bar{\tilde{m}}= \ln N \big( -2 \kappa_N^{(0)} - \kappa_N^{(1)}
+2 \kappa_N^{(2\alpha)}+2 \kappa_N^{(2\beta)} \big) \,,
\nn\\
\text{Odd $N$:}\quad &
  \bar{m}= \ln N \big( -4 \kappa_N^{(0)} -3 \kappa_N^{(1)}
-4 \kappa_N^{(2\alpha)} \big) \,,
\nn\\ &
 \bar{\tilde{m}}= \ln N \big( -2 \kappa_N^{(0)} -3 \kappa_N^{(1)}
-2 \kappa_N^{(2\alpha)}-6 \kappa_N^{(2\beta)} \big) \,.
\end{align}
Thus, we observe that the the mass parameters have different
coupling dependence near Gaussian fixed point for even and odd $N$
cases.
This suggests that even near Gaussian fixed point the critical behavior
depends on evenness/oddness of $N$.
Note that $-4 \ln N$ coefficient of $\kappa_N^{(0)}$ for $\bar{m}$
is also obtained
in the previous study \cite{Kawamoto:2012ng}, and we reproduce that result
by setting other coupling constants to be zero, for either even or odd $N$.

In order to compare the result with the one from the following 2-step study,
we present eigenvectors of this linearized transformation in
$(m^2, \tilde{m}^2, \kappa^{(0)}, \kappa^{(1)}, \kappa^{(2\alpha)}, \kappa^{(2\beta)})$
basis.
On this basis, the transformation matrix is of upper triangular form,
with sextuple degenerate eigenvalue that corresponds to the canonical scaling dimension 2, and there are four  eigenvectors,
\begin{align}
&  (1,0,0,0,0,0) \,,
\quad
(0,1,0,0,0,0) \,,
\nn\\
&
\big( 0,0,5+4(-1)^{N-1}, -4-8(-1)^{N-1}, 0,3 \big) \,,
\quad
\big( 0,0,-1-2(-1)^{N-1}, 2-2(-1)^{N-1}, 3,0 \big) \,,
\end{align}
where we have not normalized them.
Unlike the cases of the following subsections,
two of the eigenvectors are different for even and odd $N$.

\subsection{Fixed points for 2-step RGEs}
\label{sec:fixed-points-2}

The observation in the previous subsection may imply that
Wilsonian RG is not implemented on a fuzzy sphere, as Vaidya suggests in \cite{Vaidya:2001bt}.
However, as mentioned in Introduction, Chu et al. \cite{Chu:2001xi} claim that 
the singular behavior due to the oscillating phase is not a true problem,
and by carrying out loop integrals for all momenta the two point function does not have 
such an oscillating behavior.
This suggests that the problem occurs since we consider only 1-step RG transformation,
where only $l=2L$ modes are integrated out.
This motivates us to consider iterative application of RG transformation,
which leads to theory of $N \times N$ matrices to that of $(N-\hat{n})\times (N-\hat{n})$
matrices, with $\hat{n}>1$.
Choosing even $\hat{n}$ is also plausible, since evenness and oddness of $N$ is preserved
by RG transformation.
In this subsection, we consider the first nontrivial $\hat{n}=2$ case, which we call 2-step RG transformation.  
Its fixed points are expected to correspond to the large-$N$ limit taken in such a way that 
$N=2M$ or $N=2M+1$ as $M\rightarrow\infty$. 

\subsubsection{2-step RG equations}
\label{sec:2-step-rg}

By using the RG transformation \eqref{eq:m_1-step}--\eqref{eq:kappa_a_1-step}
twice, we can write down 2-step RG equations
\begin{align}
     m_{N-2}^2 =&
b_N^2 b_{N-1}^2 m_N^2
\nn\\&
+b_N^2b_{N-1}^2 \bigg(
 {B}_1(N; m_N^2, \tilde{m}_N^2) \mathcal{X}^{(1)}_N(\kappa_N)
+ \tilde{B}_1(N-1; m_N^2, \tilde{m}_N^2) \mathcal{X}^{(1)}_{N-1} (\kappa_N)
\bigg)
\nn\\&
+\mathcal{O}(\kappa^2)
\label{eq:m_2-step}
\,,\\
   \tilde{m}_{N-2}^2 =&
b_N^2 b_{N-1}^2 \tilde{m}_N^2
\nn\\&
+b_N^2b_{N-1}^2 \bigg(
 {B}_1(N; m_N^2, \tilde{m}_N^2) \mathcal{X}^{(2)}_N(\kappa_N)
+ \tilde{B}_1(N-1; m_N^2, \tilde{m}_N^2) \mathcal{X}^{(2)}_{N-1} (\kappa_N)
\bigg)
\nn\\&
+\mathcal{O}(\kappa^2)
\label{eq:mt_2-step}
\,,\\
\kappa_{N-2}^{(a)} =&
b_N^2 b_{N-1}^2 \kappa_N^{(a)}
\nn\\&
-b_N^2b_{N-1}^2 \rho_N^2
\bigg( {B}_2(N; m_N^2, \tilde{m}_N^2) \mathcal{Y}^{(a)}_N (\kappa_N)
+\tilde{B}_2(N-1; m_N^2, \tilde{m}_N^2) \mathcal{Y}^{(a)}_{N-1} (\kappa_N)
\bigg)
\nn\\&
+\mathcal{O}(\kappa^3)
\label{eq:kappa_a_2-step}
\,,
\end{align}
where we consider up to $\mathcal{O}(\kappa)$ terms for the mass parameters.
When we look for a fixed point at the leading order in $1/N$ expansion, 
we check that all $\kappa_N^{(a)}$ there are small and that 
$\mathcal{O}(\kappa^2)$ terms provide merely small corrections.
In a later subsection, we carry out linearized analysis around fixed points,
and there it is necessary to include only the leading corrections in the coupling
constants, and then the above set of equations are sufficient for our purpose.
Here, the mass parameters and the coupling constants on the right hand sides
are $m_N^2, \tilde{m}_N^2$, and $\kappa_N^{(a)}$.
If we apply our RGE twice straightforwardly, we will have $m_{N-1}^2$ and $\tilde{m}_{N-1}^2$ 
on the right hand sides, but they are all replaced with $m_N^2$ and $\tilde{m}_N^2$ respectively 
by using the RGEs again. More precisely, in the second application of the RGEs, 
there appear propagators with $m_{N-1}^2$ and $\tilde{m}_{N-1}^2$, namely  
$B_1(N-1; m_{N-1}^2, \tilde{m}_{N-1}^2)$ and $B_2(N-1; m_{N-1}^2, \tilde{m}_{N-1}^2)$.
These mass parameters are to be further replaced by the RGEs 
\begin{align}
  m_{N-1}^2 =& b_N^2 m_N^2 + \mathcal{O}(\kappa) \,,
\qquad
  \tilde{m}_{N-1}^2 = b_N^2 \tilde{m}_N^2 + \mathcal{O}(\kappa) \,,
\end{align}
but, to the order we take now, it is sufficient to take the leading terms.
This leads to
\begin{align}
  B_1(N-1; m_{N-1}^2, \tilde{m}_{N-1}^2) =&
\tilde{B}_1(N-1;m_{N}^2, \tilde{m}_{N}^2) + \mathcal{O}(\kappa) \,,
\nn\\
\tilde{B}_1(N-1;m_{N}^2, \tilde{m}_{N}^2) =& 2(2N-3) \tilde{P}_{N-1} \,,
\nn\\
\tilde{P}_{N-1} =& \frac{1}{(N-1)(N-2) + \rho_N^2(m_N^2 + (-1)^{N-2}\tilde{m}_N^2)} \,,
\end{align}
and $\tilde{B}_2(N-1;m_{N}^2, \tilde{m}_{N}^2)=2(2N-3)\tilde{P}_{N-1}^2$.
Note that we have used the relation $b_{N}^2 \rho_{N-1}^2= \rho_N^2$. 
In eqs.~\eqref{eq:m_2-step}--\eqref{eq:kappa_a_2-step}, 
$\mathcal{X}^{(i)}_{N-1}(\kappa_N)$ ($i=1,2$) and 
$\mathcal{Y}^{(a)}_{N-1} (\kappa_N)$ 
mean that the $N$ dependent coefficient $(-1)^{N-1}$ in \eqref{X1}--\eqref{Y2b} 
should be replaced with $(-1)^{N-2}$ but the arguments are still $\kappa_N^{(a)}$. 
Namely, the subscript refers to the $N$ dependence of the functions $\mathcal{X}_{N-1}^{(i)}(\cdot)$ 
and $\mathcal{Y}_{N-1}^{(a)}(\cdot)$ themselves.
Thus, on these RGEs, the mass parameter dependence appears through
$B_i$ coefficients, while the coupling constants dependence is from $\mathcal{X}$
and $\mathcal{Y}$.

Before going to fixed point analysis,
we take a look at the validity of 2-step RGEs just presented.
Here, a map from $N\times N$ theory to $(N-2)\times (N-2)$ theory
is defined by applying 1-step RG transformation twice,
which can in principle be different from the one defined by
integrating out both $l=2L$ and $l=2L-1$ modes at the same time, 
if we make some approximations. 
For example, one possible difference is from a graph involving two out-mode propagators,
where one is $l=2L$ mode and the other is $l=2L-1$. 
The latter does include these contributions, while the former (the ones we have just presented) 
does not due to the low energy approximation used to derive the RG transformation. 
The reason is that in the first step RG integrating only $l=2L$, all the other lines are assumed 
to have sufficiently low angular momentum $l\ll L$ by the low energy approximation, 
but in order to reproduce this graph one of them should have $l=2L-1$ in the next step RG, 
and hence the low energy approximation assumed in the first step does not work. 
Such difference can be understood as potential error terms for iterative application
of RG transformation with low energy approximation.
It is therefore important to estimate such difference.

In Appendix \ref{app:calc_VEVs}, we present formulas for expectation values
by integrating $l=2L, \cdots, 2L-\hat{n}+1$ out modes.\footnote{%
Precisely speaking, one of the formulas is only valid for $\hat{n}=2$ case,
due to a technical difficulty (we are about to mention it).
But it is sufficient for the current purpose.}
We use them to derive the RGEs from $N\times N$ theory to $(N-2)\times (N-2)$
theory by integrating out $l=2L, 2L-1$ modes.
We do not present the details in this paper, but one can follow the calculation
by use of the formulas in Appendix \ref{app:calc_VEVs}.
We have found that, to $\mathcal{O}(\kappa)$ for the masses and to $\mathcal{O}(\kappa^2)$
for the coupling constants,
the result is exactly the same as \eqref{eq:m_2-step}--\eqref{eq:kappa_a_2-step}.
The difference appears in $\mathcal{O}(\kappa^2)$ terms for the mass RGEs
\eqref{eq:m_2-step} and \eqref{eq:mt_2-step}.
Therefore, we can trust our RGEs for $\hat{n}=2$ case.
Interestingly, this observation also suggests that we may formulate $\hat{n}$-step
RGEs just by applying the 1-step RGE $\hat{n}$ times.
To the leading order in $1/N$ expansion, the difference can appear only in
$\mathcal{O}(\kappa^2)$ terms in the mass RGEs
and $\mathcal{O}(\kappa^3)$ terms in the coupling constant ones.
We cannot fully justify this observation for now, as we do not have a complete formula
for $\vev{\mathcal{V}_2^{(a)} \mathcal{V}_2^{(b)}}_c$ expectation value, but the
structure of the asymptotic formula for $9j$ symbols seems to support this conjecture.
If this is the case, then we can carry out RG analysis to $(N-\hat{n})\times (N-\hat{n})$
size, where $\hat{n} \ll N$ but can be very large.
This completes the large-$N$ RG analysis and, for example, enables us to draw
the global picture of RG flows. 
However, we will leave this
for future study, and
now concentrate on $\hat{n}=2$ case.
From the following subsection, we look for fixed points and study their properties.

\subsubsection{Fixed points}
\label{sec:fixed-points}

As before, we set
\begin{align}
  m_{N-2}^2 = m_N^2 = m_*^2 \,,
\quad
  \tilde{m}_{N-2}^2 =  \tilde{m}_N^2 = \tilde{m}_*^2 \,,
\quad
\kappa_{N-2}^{(a)} = \kappa_N^{(a)}=\kappa_*^{(a)} \,,
\end{align}
and solve the following six equations obtained 
from \eqref{eq:m_2-step}--\eqref{eq:kappa_a_2-step},
\begin{align}
     m_*^2 =&
b_N^2 b_{N-1}^2 m_*^2
+b_N^2b_{N-1}^2 \bigg(
 {B}_1^*(N) \mathcal{X}^{(1)}_N(\kappa_*)
+ \tilde{B}_1^*(N-1) \mathcal{X}^{(1)}_{N-1} (\kappa_*)
\bigg)
\label{eq:m*_2-step}
\,,\\
   \tilde{m}_*^2 =&
b_N^2 b_{N-1}^2 \tilde{m}_*^2
+b_N^2b_{N-1}^2 \bigg(
 {B}_1^*(N) \mathcal{X}^{(2)}_N(\kappa_*)
+ \tilde{B}_1^*(N-1) \mathcal{X}^{(2)}_{N-1} (\kappa_*)
\bigg)
\label{eq:mt*_2-step}
\,,\\
\kappa_*^{(a)} =&
b_N^2 b_{N-1}^2 \kappa_*^{(a)}
-b_N^2b_{N-1}^2 \rho_N^2
\bigg( {B}_2^*(N) \mathcal{Y}^{(a)}_N (\kappa_*)
+\tilde{B}_2^*(N-1) \mathcal{Y}^{(a)}_{N-1} (\kappa_*)
\bigg)
\label{eq:kappa_a*_2-step}
\,,
\end{align}
where
\begin{align}
{B}_1^*(N)=&  {B}_1(N; m_*^2, \tilde{m}_*^2) = 2(2N-1) P_N^* \,,
 \quad
\tilde{B}_1^*(N-1)
 = 2(2N-3) \tilde{P}_{N-1}^* \,,
\nn\\
{B}_2^*(N)=& 
2(2N-1) {P_N^*}^2 \,,
\quad
\tilde{B}_2^*(N-1)=
2(2N-3) \tilde{P}_{N-1}^{*\,2} \,,
\nn\\
P_N^* =& \frac{1}{N(N-1) + \rho_N^2 (m_*^2 + (-1)^{N-1}\tilde{m}_*^2)} \,,
\nn\\
\tilde{P}_{N-1}^* =& \frac{1}{(N-1)(N-2) + \rho_N^2 (m_*^2 - (-1)^{N-1}\tilde{m}_*^2)} \,.
\end{align}
We introduce the following rescaled variables,
\begin{align}
  m_*^2 =&  \frac{x_*^{(1)}}{\rho_N^2 P_N^*} \,,\qquad
  \tilde{m}_*^2 = \frac{x_*^{(2)}}{\rho_N^2 P_N^*} \,,\qquad
\kappa_*^{(a)}= \frac{b_N^2b_{N-1}^2-1}{b_N^2b_{N-1}^2 \rho_N^2 B_2^*(N)} y_*^{(a)}
\,.
\label{def:rescaled_2step}
\end{align}
$P_N^*$ and $\tilde{P}_{N-1}^*$ can be written in terms of $x_*^{(i)}$ ($i=1,2$) as
\begin{align}
  {P}_N^* =&
\frac{1-\big( x_*^{(1)} + (-1)^{N-1} x_*^{(2)} \big) }{N(N-1)}
\,,
\label{eq:P*}
\\
\tilde{P}_{N-1}^* =&
\frac{1}{N(N-1)}\frac{1-(x_*^{(1)} + (-1)^{N-1} x_*^{(2)})}
{1-2(-1)^{N-1}x_*^{(2)}-\frac{2}{N}\big(1-(x_*^{(1)} + (-1)^{N-1} x_*^{(2)})\big)}
\,.
\label{eq:P*'}
\end{align}
With these variables, the fixed point equations are
\begin{align}
0=& x_*^{(i)} + \mathcal{X}_N^{(i)}(y_*) + \frac{\tilde{B}_1^*(N-1)}{B_1^*(N)} \mathcal{X}_{N-1}^{(i)}(y_*) \,, \nn\\
  0=& y_*^{(a)} - \mathcal{Y}_N^{(a)}(y_*) - \frac{\tilde{B}_2^*(N-1)}{B_2^*(N)}
\mathcal{Y}_{N-1}^{(a)}(y_*) \,.
\label{fixed_point_eq_2step}
\end{align}
They are again algebraic equations with numerical coefficients (for a given $N$).
We assume that fixed points $x_*^{(i)}$ and $y_*^{(a)}$ are of order 1,
and then take the leading order coefficients in the large-$N$ limit.
Since $\frac{\tilde{B}_2^*(N-1)}{B_2^*(N)}$ depends on $x_*^{(i)}$, these six equations are now coupled, but again we can numerically find four real
solutions for even and odd $N$ respectively.

For even $N$, we find
\begin{align}
  \big( x_*^{(1)}, x_*^{(2)}, y_*^{(0)}, y_*^{(1)}, y_*^{(2\alpha)}, y_*^{(2\beta)} \big) =&
\big(0,0,0,0,0,0 \big) \,,\nn\\
& \big( -0.1671,\, 0.1671,\, 0.1902,\, -0.7608,\, 0.09582,\, 0.4748 \big)\,,
 \nn\\ 
& \big( -0.06894,\, 0.1267,\, 0.08281,\, -0.6088,\, 0.1656,\, 0.4766 \big)\,, \nn\\ 
& \big( -0.1198,\, -0.1270,\, 0.01456,\, 0.06177,\, 0.02912,\, 0.01820 \big)\,.
\label{xyfpeven}
\end{align}
Let us call these four solutions $\mathfrak{e}(i)$ ($i=0,1,2,3$) in order.
On the other hand, for odd $N$,
\begin{align}
  \big( x_*^{(1)}, x_*^{(2)}, y_*^{(0)}, y_*^{(1)}, y_*^{(2\alpha)}, y_*^{(2\beta)} \big) =&
\big( 0,0,0,0,0,0 \big) \,,\nn\\
&
\big( -0.1252,\, 0.1252,\, 0.1068,\, -0.4274,\, 0.05383,\, 0.2667 \big)\,,
 \nn\\
&
\big( -0.05500,\, 0.1011,\, 0.05271,\, -0.3875,\, 0.1054,\, 0.3034 \big) \,,
\nn\\ 
&
\big( -0.1606,\, -0.1703,\, 0.02616,\, 0.1110,\, 0.05233,\, 0.03271 \big) 
\,,
\label{xyfpodd}
 \end{align}
and we call them $\mathfrak{o}(i)$ ($i=0,1,2,3$). Apart from the Gaussian fixed point case, 
the numerical computation is done with 15 digits working precision,
but we round the results to four significant figures and show them.
So far, we have numerically confirmed these sets of solutions
and it seems very unlikely that there exists another.

At a glance, they do not seem to share common solutions.
However, the physical quantities are the original parameters
$m_N^2$ and the others.
The rescaling factors in \eqref{def:rescaled_2step}
also depends on $N$ as in \eqref{eq:rhoandalpha}, \eqref{eq:P*} 
and thus we indeed find that the fixed point values of the masses and the coupling constants are
the same for both $\mathfrak{e}(i)$ and $\mathfrak{o}(i)$ given by 
\begin{align}
&\big( \alpha^2m_*^2 , \alpha^2\tilde m_*^2, \alpha^2\kappa_*^{(0)}, 
\alpha^2\kappa_*^{(1)}, \alpha^2\kappa_*^{(2\alpha)}, \alpha^2\kappa_*^{(2\beta)} \big) \nn \\ 
=&
\big(
0, 0, 0, 0, 0, 0  \big) 
\,,\nn\\
&
\big(
-0.5010, 0.5010, 0.4274, -1.710, 0.2153, 1.067 
\big) \,,\nn\\
&
\big(
-0.2306, 0.4239, 0.2317, -1.703, 0.4634, 1.334 
\big)\,,\nn\\
&
\big(
-0.4826, -0.5119, 0.05909, 0.2507, 0.1182, 0.07387 
\big)
\,,
\label{eq:fpts_2-step}
\end{align}
for $i=0,1,2,3$ in order. 
Here we have neglected subleading contributions in the $1/N$ expansion 
and $\alpha$ is the scale fixed in our RG describing the fuzzy sphere limit 
given in \eqref{eq:rhoandalpha}. 
Therefore, our 2-step RGE turns out to have the common fixed points
for even and odd $N$ cases. 
It seems nontrivial because the fixed point equations 
\eqref{eq:m*_2-step}--\eqref{eq:kappa_a*_2-step} change for even and odd $N$. 
However one would anticipate agreement of the fixed points, 
since the fixed point action which describes the large-$N$ limit 
of our matrix model should not depend on evenness or oddness of $N$.
Actually, with hindsight, we may have foreseen
this agreement from the fixed point
equations \eqref{eq:m_2-step}--\eqref{eq:kappa_a_2-step} in the large-$N$ limit.
For example, we consider the leading order part of $1/N$ in \eqref{eq:m_2-step},
with $\rho_N^2$, $b_N^2b_{N-1}^2$, and the phases left untouched,
\begin{align*}
&
\frac{N}{4 b_N^2b_{N-1}^2}\big(1-b_N^2 b_{N-1}^2 \big)     m_*^2
\nn\\ =&
 \frac{1}{1+\frac{\rho_N^2}{N^2}(m_*^2 + (-1)^{N-1}\tilde{m}_*^2)}
\mathcal{X}^{(1)}_N(\kappa_*)
+\frac{1}{1+\frac{\rho_N^2}{N^2}(m_*^2 + (-1)^{N-2}\tilde{m}_*^2)}
 \mathcal{X}^{(1)}_{N-1} (\kappa_*)
  \,.
\end{align*}
On the right hand side, the phase factors in the first term are all $(-1)^{N-1}$,
including the ones in $\mathcal{X}^{(1)}_N(\kappa_*)$,
while $(-1)^{N-2}$ for the second term.
Thus the right hand side takes the same form for even and odd $N$.
One can easily check that this is true for the other fixed point equations.
Therefore, the fixed points in terms of the original variables should be
the same for even and odd N.
Note that this argument is more or less formal;
 the parameters have nontrivial $N$ dependence
of which we need to take care when we solve the fixed point equations.
We have just observed that the (leading order) 
structure of the fixed point equations admits
common solutions for even and odd $N$.
At any rate,
the fact that the even/odd $N$ have the same fixed points strongly supports 
validity of our RG with antipode fields.

Furthermore, from the field theory point of view, our result also seems to be consistent 
with the claim in \cite{Chu:2001xi}: 
it is true that when we integrate only the highest mode, 
the result would not be a smooth function with different values for even and odd $N$ 
as pointed out in \cite{Vaidya:2001bt}, but integrating all modes in calculating correlation functions 
makes them smooth and completely regular in the $1/N$ expansion. 
In our large-$N$ RG, we have only considered the integration over the highest modes 
and hence the RG itself is not smooth in the sense that it has the explicit oscillating phase $(-1)^{N-1}$ 
as found in \cite{Vaidya:2001bt}. The point here is that integration over all modes is realized 
in our approach by looking at the fixed points. There the phase dependence in fact disappears and 
we have the well-defined fixed points. Note that the large-$N$ limit, more precisely the fuzzy sphere limit 
we are considering is exactly the same as in \cite{Chu:2001xi}. 

We also notice that since the loop expansion parameter is in general 
$\rho_N^2\kappa_N^{(a)}P_N^2N$ as seen in \eqref{eq:m_1-step}--\eqref{eq:kappa_a_1-step}, 
the finite fixed points in \eqref{eq:fpts_2-step} are consistent with our perturbative approach of the RG. 
In the following subsection, we further investigate the scaling dimensions
of the operators associated with each fixed point, which also imply 
that our RG is legitimate enough to capture universality in the large-$N$ limit.

\subsubsection{Linearized analysis and scaling dimensions}
\label{sec:line-analys-crit}

In this subsection, we carry out linearized analysis around
the fixed points found in the previous subsection.

We start with $m_N^2$ RGE.
By subtracting \eqref{eq:m*_2-step} from \eqref{eq:m_2-step},
and defining $\delta m_{N-2}^2= m_{N-2}^2 - m_*^2$, 
$\delta m_{N}^2= m_{N}^2 - m_*^2$ and so on,
we find, up to the linear order in $\delta m_{N}^2$ and the others,
\begin{align}
  \delta m_{N-2}^2 =&
b_N^2 b_{N-1}^2 \delta m_N^2
\nn\\&
+ b_N^2 b_{N-1}^2 \bigg[\delta m_N^2 \partial_{m_N^2}
+  \delta {\tilde{m}_N^2} \partial_{\tilde{m}_N^2} 
+\sum_a  \delta {\kappa_N^{(a)}} \partial_{\kappa_N^{(a)}}
\bigg]
\nn\\& \hskip1em \times
 \bigg(
 {B}_1(N; m_N^2, \tilde{m}_N^2) \mathcal{X}^{(1)}_N(\kappa_N)
+ \tilde{B}_1(N-1; m_N^2, \tilde{m}_N^2) \mathcal{X}^{(1)}_{N-1} (\kappa_N)
\bigg) \bigg|_\text{fixed point}
\end{align}
where $m_N^2$, $\tilde{m}_N^2$, and $\kappa_N^{(a)}$ on the right hand
side will be replaced with their fixed point values after taking
the derivatives.
By noting that the mass derivatives only act on $B_1$, $\tilde{B}_1$ 
and the $\kappa_N^{(a)}$ derivatives on $\mathcal{X}$,
we find from \eqref{eq:P*B1*B2*}
\begin{align}
  \delta m_{N-2}^2 =&
b_N^2 b_{N-1}^2 \delta m_N^2
\nn\\&
+b_N^2 b_{N-1}^2 B_2^*(N) \mathcal{X}^{(1)}_N(\kappa_*)
(-\rho_N^2) \big[ \delta m_N^2 + (-1)^{N-1} \delta \tilde{m}_N^2 \big]
\nn\\&
+b_N^2 b_{N-1}^2 \tilde{B}_2^*(N-1) \mathcal{X}^{(1)}_{N-1}(\kappa_*)
(-\rho_N^2) \big[ \delta m_N^2 - (-1)^{N-1} \delta \tilde{m}_N^2 \big]
\nn\\&
+b_N^2 b_{N-1}^2 \sum_a  \delta {\kappa_N^{(a)}} 
\bigg[ B_1^*(N) \partial_{\kappa_*^{(a)}} \mathcal{X}^{(1)}_N(\kappa_*)
+\tilde{B}_1^*(N-1) \partial_{\kappa_*^{(a)}} \mathcal{X}^{(1)}_{N-1}(\kappa_*)
\bigg]
\nn\\=&
b_N^2 b_{N-1}^2 \delta m_N^2
\nn\\&
+\frac{1-b_N^2 b_{N-1}^2 }{B_2^*(N)}\big[
B_2^*(N) \mathcal{X}^{(1)}_N(y_*) + \tilde{B}_2^*(N-1)\mathcal{X}^{(1)}_{N-1}(y_*)
\big] \delta m_N^{2}
\nn\\&
+(-1)^{N-1} \frac{1-b_N^2 b_{N-1}^2 }{B_2^*(N)}\big[
B_2^*(N) \mathcal{X}^{(1)}_N(y_*) - \tilde{B}_2^*(N-1)\mathcal{X}^{(1)}_{N-1}(y_*)
\big] \delta \tilde{m}_N^{2}
\nn\\&
+b_N^2 b_{N-1}^2
\sum_a  \big[ B_1^*(N) \partial_{y_*^{(a)}} \mathcal{X}^{(1)}_N(y_*)
+ B_1^*(N-1)  \partial_{y_*^{(a)}} \mathcal{X}^{(1)}_{N-1}(y_*) \big] \delta \kappa_N^{(a)}
\label{eq:lin_RGE_mass1}
\,,
\end{align}
where in the final part, we have noted that $\mathcal{X}^{(1)}_N(\kappa_*)$ 
and $ \mathcal{X}^{(1)}_{N-1}(\kappa_*)$ are linear in $\kappa_*^{(a)}$ 
and used the rescaled variables \eqref{def:rescaled_2step} to express the coefficients
by use of the values of the fixed points we have just found.
It should also be noted that since
the rescaling factors depends on $m_*^2$
and $\tilde{m}_*^2$
but not on $\kappa_*^{(a)}$, $\kappa_*^{(a)}$ derivative
does not act on the rescaling factors.
In the similar way, one can evaluate $\delta \tilde{m}_{N-2}$ and
$\delta \kappa_{N-2}^{(a)}$, and the result is summarized in the following
matrix form,
\begin{align}
  \delta \bm{v}_{N-2} =& \big( b_N^2 b_{N-1}^2 \mathbf{1}_6
+ \tilde{M} \big) \delta \bm{v}_N \,,
\label{lin_RGE1}
\end{align}
where $\delta \bm{v}_N = ( \delta m_N^2, \delta \tilde{m}_N^2, \delta \kappa_N^{(0)},
\delta \kappa_N^{(1)},\delta \kappa_N^{(2\alpha)},\delta \kappa_N^{(2\beta)})^T$,
and $\delta \bm{v}_{N-2}$ is defined in the similar way.
$\mathbf{1}_6$ is $6 \times 6$ unit matrix and $\tilde{M}$ is a $6\times 6$ matrix whose
elements are from the derivative part of the previous relations,
\begin{align}
  \tilde{M}_{i1}=& 
\frac{1-b_N^2 b_{N-1}^2 }{B_2^*(N)}\big[
B_2^*(N) \mathcal{X}^{(i)}_N(y_*) + \tilde{B}_2^*(N-1)\mathcal{X}^{(i)}_{N-1}(y_*)
\big] \,,
\nn\\
  \tilde{M}_{i2}=&  (-1)^{N-1}
\frac{1-b_N^2 b_{N-1}^2 }{B_2^*(N)}\big[
B_2^*(N) \mathcal{X}^{(i)}_N(y_*) - \tilde{B}_2^*(N-1)\mathcal{X}^{(i)}_{N-1}(y_*)
\big] \,,
\nn\\
  \tilde{M}_{ia}=&
b_N^2 b_{N-1}^2
 \big[ B_1^*(N) \partial_{y_*^{(a)}} \mathcal{X}^{(i)}_N(y_*)
+ B_1^*(N-1)  \partial_{y_*^{(a)}} \mathcal{X}^{(i)}_{N-1}(y_*) \big]
\,,\nn\\
\tilde{M}_{a1}=&
\frac{2(b_N^2 b_{N-1}^2-1)^2}{b_N^2 b_{N-1}^2 [B_2^*(N)]^2}
\bigg(B_3^*(N) \mathcal{Y}^{(a)}_N (y_*)
+\tilde{B}_3^*(N-1) \mathcal{Y}^{(a)}_{N-1} (y_*)
\bigg) \,,
\nn\\
\tilde{M}_{a2}=&
(-1)^{N-1} \frac{2(b_N^2 b_{N-1}^2-1)^2}{b_N^2 b_{N-1}^2 [B_2^*(N)]^2}
\bigg(B_3^*(N) \mathcal{Y}^{(a)}_N (y_*)
-\tilde{B}_3^*(N-1) \mathcal{Y}^{(a)}_{N-1} (y_*)
\bigg) \,,
\nn\\
\tilde{M}_{ab}=&
\frac{1- b_N^2 b_{N-1}^2 }{B_2^*(N)}
\bigg(B_2^*(N) \partial_{y_*^{(b)}} \mathcal{Y}^{(a)}_N (y_*)
+\tilde{B}_2^*(N-1) \partial_{y_*^{(b)}} \mathcal{Y}^{(a)}_{N-1} (y_*)
\bigg)
\,,
\label{def:tilde_M}
\end{align}
where $i=1,2$. 
On the left hand side, the ordered set $(i,a)$ is understood to label
the indices of the matrix.
We have introduced $B_3^*(N)=B_2^*(N)P_N^*$ and $\tilde{B}_3(N-1) = \tilde{B}_2(N-1)\tilde{P}_{N-1}^*$ for simplicity.
By recalling that the $N$ dependence of each coefficient factor,
\begin{align}
  b_N^2 b_{N-1}^2 = 1+ \mathcal{O}(N^{-1})\,,
\qquad
B_k^*(N) = \tilde{B}_k^* (N-1) = \mathcal{O}(N^{1-2k}) \quad (k=1,2,3) \,,
\label{eq:Border}
\end{align}
and the fact that the values of the fixed points are of order 1,
one can see that all the elements of $\tilde{M}$ is of $\mathcal{O}(N^{-1})$.
We are going to find the eigenvalues $\lambda$ for this RG transformation matrix
$b_N^2 b_{N-1}^2 \mathbf{1}_6 + \tilde{M}$.

Before going, we make a remark on the eigenvalues of the above linearized transformation matrix
and the scaling dimensions associated with a given fixed point.
Suppose that there exists an operator $\mathcal{O}_\Delta$ of scaling dimension $\Delta$ at a fixed point.
Near the fixed point, the operator receives a scale transformation through the RG transformation,
\begin{align}
  \mathcal{O}_\Delta \rightarrow b_N^\Delta \mathcal{O}_\Delta \,,
\end{align}
where $b_N=\rho_N/\rho_{N-1}$ is the scale factor associated with the RG transformation defined 
in \eqref{scale}.
As we have discussed in Section \ref{sec:rescaling}, $b_N$ in general has the following
form
\begin{align}
  b_N^2 =& 1 + \frac{\gamma}{N} + \mathcal{O}(N^{-2}) \,,
\end{align}
where $\gamma=2$ for the fuzzy sphere limit and $\gamma=1$ for the NCFT limit.
In the current consideration, we treat a sequence of the transformations,
$N \rightarrow N-1 \rightarrow N-2$, and then the $\mathcal{O}_\Delta$ would have
the following factor under the linearized transformation matrix,
\begin{align}
  b_N^\Delta b_{N-1}^\Delta =& 1 + \frac{\gamma \Delta}{N} + \mathcal{O}(N^{-2}) \,.
\label{eq:evofRG}
\end{align}
On the other hand, since our RG transformation only changes $N\rightarrow N-2$ in total, 
the leading of the eigenvalue of the linearized RG should be 1. Moreover, \eqref{eq:evofRG} 
implies that the eigenvalue in general takes a form $\lambda=1+w/N+\mathcal{O}(1/N^2)$ 
(i.e. no fractional power of $N$) in the $1/N$ expansion 
from which we find that an operator corresponding to an eigenvector 
belonging to $\lambda$ has the scaling dimension $\Delta=w/\gamma$.  
More concretely, \eqref{lin_RGE1} gives an explicit form of the linearized RG in the $1/N$ expansion 
\begin{align}
  b_N^2 b_{N-1}^2 \mathbf{1}_6 + \tilde{M}
= \mathbf{1}_6 + \frac{1}{N} M + \mathcal{O}(N^{-2}) \,,
\label{eq:linearizedRG}
\end{align}
where
\begin{align}
  M=& 2\gamma \mathbf{1}_6 + \tilde{M}\big|_{\mathcal{O}(N^{-1})} \,,
\end{align}
and $\tilde{M}\big|_{\mathcal{O}(N^{-1})}$ means that we take the leading order term 
(namely, that of ${\mathcal O}(1/N)$ as mentioned below \eqref{eq:Border})  for each matrix element.
From this form, it is manifest that if $\mathcal{O}(1/N)$ terms are dropped,
the eigenvalue of the RG transformation matrix is 1, the sextuple root. 
Thus if $\bm v$ is an eigenvector belonging to an eigenvalue $\lambda=1+w/N$ of 
the linearized RG in \eqref{eq:linearizedRG}, it follows that 
\begin{align}
M\bm v=w\bm v,
\end{align}
up to ${\cal O}(1/N)$ corrections, 
and that $\bm v$ corresponds to an operator with scaling dimension $\Delta=w/\gamma$. 
The general argument above predicts that $w$ is $\mathcal{O}(1)$ and this is indeed the case 
because $M$ is a matrix of $\mathcal{O}(1)$. 
Note that when the loop corrections are neglected, which corresponds to dropping $\tilde{M}$,
the eigenvalue of $M$ is $2\gamma$ (sextuple root), and the scaling dimensions are all
$\Delta=2$, as anticipated.

Now, we can calculate the eigenvalue and the eigenvectors associated with
each fixed point for $N$ even and odd respectively.
For numerical study, we also need to fix a value of $\gamma$ from $b_N^2$
defined above.
Thus, we study $\gamma=2$ case (the fuzzy sphere limit) and $\gamma=1$ case
(the NCFT limit) separately.
It has been found that after substituting the values of the fixed points,
the matrix elements \eqref{def:tilde_M} of even and odd $N$ coincides,
up to very small numerical errors, for suitable pairs of fixed point values.
Therefore, in this 2-step RGE case, even and odd $N$ cases
share the same scaling dimensions (critical exponents)
and the corresponding eigenvectors (namely, associated operators)
are also the same.
As in the case of the fixed point values,
we can see that this agreement comes from the structure of the linearized
RGE.
For example, from the middle expression in \eqref{eq:lin_RGE_mass1},
the coefficient of $\delta m_N^2$ (namely the matrix element $\tilde{M}_{11}$)
reads
\begin{align*}
-\rho_N^2  b_N^2 b_{N-1}^2
\big[  B_2^*(N) \mathcal{X}^{(1)}_N(\kappa_*)
+ \tilde{B}_2^*(N-1) \mathcal{X}^{(1)}_{N-1}(\kappa_*) \big]\,.
\end{align*}
If we consider the leading order part of $1/N$
with $\rho_N^2$, $b_N^2b_{N-1}^2$, and the phases left untouched,
it becomes
\begin{align*}
-\frac{4\rho_N^2  b_N^2 b_{N-1}^2}{N^3}
\bigg[
\frac{\mathcal{X}^{(1)}_N(\kappa_*)}{[1+\frac{\rho_N^2}{N^2}(m_*^2 + (-1)^{N-1}\tilde{m}_*^2)]^2}
+\frac{\mathcal{X}^{(1)}_{N-1}(\kappa_*)}{[1+\frac{\rho_N^2}{N^2}(m_*^2 + (-1)^{N-2}\tilde{m}_*^2)]^2}
  \bigg]\,.
\end{align*}
Again, the phase factors in the first term in the square bracket
are all $(-1)^{N-1}$, including the ones in $\mathcal{X}^{(1)}_N(\kappa_*)$,
while $(-1)^{N-2}$ for the second term.
Thus, if the fixed point values of $\kappa_*^{(a)}$ are
the same for even and odd $N$, which has already been confirmed,
the matrix element $\tilde{M}_{11}$ is the same for
even and odd $N$.
One can check that this is also true for
all the other matrix elements.
Thus,
we can conclude that even and odd $N$ cases share the common eigenvalues
and eigenvectors at the leading order in the large-$N$ limit,
and this will be true for any choice of $\gamma$
as long as fixed point values in terms of the physical variables
coincide.

The expression \eqref{def:tilde_M} is suitable for numerical study.
We present the eigenvalues $w$ and the corresponding normalized eigenvectors 
on the basis $(\delta m^2, \delta \tilde{m}^2, \delta \kappa^{(0)},
\delta \kappa^{(1)},\delta \kappa^{(2\alpha)},\delta \kappa^{(2\beta)})$
respectively for each fixed point  
$\mathfrak{e}(i)$ (or
equivalently $\mathfrak{o}(i)$) for $i=0,1,2,3$ in order.
For the fuzzy sphere limit ($\gamma=2$), we find\footnote[1]{%
Here, we present numerical eigenvectors as row vectors.}
\begin{align}
& \mathfrak{e}(0)  (\text{Gaussian}): \nn \\
& w =4 \, (\text{sextuple})
\qquad
(1,0,0,0,0,0) \,,
\quad
(0,1,0,0,0,0) \,,
\nn\\& \hskip9em
\frac{1}{\sqrt{26}}(0,0,1,-4,0,3) \,,
\quad
\frac{1}{\sqrt{17}}(0,0,-2,2,3,0) \,,
\nn\\
\nn\\
& \mathfrak{e}(1): \nn \\
& w=4.000\, (\text{double}) \qquad \big( 0.8541 ,\; 0.1223 ,\; 0.1043 ,\; -0.4173 ,\; 0.05256 ,\; 0.2604 \big)
\nn\\ 
& w=2.884 \pm 1.548 i \qquad 
\big( -0.4987 \pm 0.01239 i ,\; 0.4987  \mp 0.01239 i ,\; -0.1406 ,\; 0.5625 ,\; \nn \\
&\hspace{8cm}-0.1511  \mp 0.1870 i ,\; -0.2708 \pm 0.1870 i \big) \nn\\ 
& w=0.9568 \qquad \big( -0.6558 ,\; -0.6078 ,\; 0.3538 ,\; 0.02579 ,\; -0.1540 ,\; -0.2257 \big) \nn\\ 
&  w=-5.308 \qquad \big( -0.1434 ,\; 0.1434 ,\; 0.2011 ,\; -0.8044 ,\; 0.09024 ,\; 0.5131 \big),
 \nn\\ 
\nn \\
& \mathfrak{e}(2): \nn \\
& w=4.000\, (\text{double}) \qquad \big( -0.8572 ,\; -0.09639 ,\; -0.05270 ,\; 0.3874 ,\; -0.1054 ,\; -0.3033 \big) \nn\\
& w=2.762 \pm 1.424 i \qquad \big( 0.3712  \mp 0.1930 i ,\; -0.6057 ,\; 0.08151 \pm 0.002246 i ,\; -0.5079 \pm 0.03613 i ,\nn \\
&\hspace{9.3cm}\; 0.2629 \pm 0.1596 i ,\; 0.2013  \mp 0.2389 i \big) \nn\\ 
& w=-1.113 \qquad \big( -0.1627 ,\; 0.6135 ,\; 0.2559 ,\; -0.2851 ,\; -0.4879 ,\; -0.4608 \big) \nn\\ 
& w=-4.687 \qquad \big( -0.06735 ,\; 0.1282 ,\; 0.1035 ,\; -0.7557 ,\; 0.1935 ,\; 0.5998 \big), 
\nn\\ 
\nn \\ 
& \mathfrak{e}(3): \nn \\
& w=4.000\, (\text{double}) \qquad \big( -0.9862 ,\; 0.1435 ,\; -0.01657 ,\; -0.07029 ,\; -0.03313 ,\; -0.02071 \big) \nn \\
& w=3.991 \qquad \big( 0.6790 ,\; 0.7213 ,\; 0.02751 ,\; 0.1169 ,\; 0.05517 ,\; 0.03445 \big) \nn\\ 
& w=3.767 \qquad \big( -0.8224 ,\; -0.5475 ,\; 0.01618 ,\; -0.1122 ,\; -0.09538 ,\; -0.04384 \big) \nn\\ 
& w=2.667 \qquad \big( 0.9526 ,\; -0.1993 ,\; -0.1283 ,\; 0.05703 ,\; 0.04414 ,\; 0.1767 \big) \nn\\ 
& w=-5.325 \qquad \big( 0.5274 ,\; 0.5594 ,\; -0.1290 ,\; -0.5474 ,\; -0.2580 ,\; -0.1613 \big)\,.
\label{evs}
\end{align}

From the discussion above, the scaling dimensions $\Delta$ are given by $\Delta=w/2$ 
for the fuzzy sphere limit case.
Thus, the scaling dimensions associated with each fixed point are
\begin{align}
&
\mathfrak{e}(1): \quad \Delta=2.000 \; (\text{double}) , \, 
1.442 \pm 0.7739 i , \, 0.4784, \, -2.654, \,  \nn\\
& \mathfrak{e}(2): \quad \Delta=2.000 \; (\text{double}) , \, 
1.381 \pm 0.7120 i ,\,  -0.5563 ,\, -2.344 ,\,   \nn\\
& \mathfrak{e}(3): \quad \Delta=2.000 \; (\text{double}) , \, 
1.996 ,\, 1.884 ,\,  1.334 ,\,     -2.662 ,\,
\label{scalingfs}
\end{align}
where $\mathfrak{e}(0)$, namely Gaussian fixed point, case always has
$\Delta=2$ as sextuple one, and is omitted from the list.
{}From \eqref{evs} and \eqref{scalingfs}, 
we obtain the following observations:
\begin{itemize}
\item There exist doubly degenerate $\Delta=2.000$ within numerical errors 
for all cases.
Numerically two eigenvectors with $\Delta=2.000$ also coincide with each other. 
The eigenvectors get significant contribution from $\delta m_N^2$ component; 
about 97$\%$ for $\mathfrak{e}(3)$ case, and 
about 73$\%$ for $\mathfrak{e}(1)$, $\mathfrak{e}(2)$. 
Thus, we can naturally identify them as a ``mass'' term around each fixed point. 
They are the most relevant operator for all fixed points. 
\item  For $\mathfrak{e}(1)$ and $\mathfrak{e}(2)$, we have complex eigenvalues 
(thus complex scaling dimensions) and eigenvectors associated with them. 
The matrix $M$ is a real matrix, and then complex eigenvalues 
and eigenvectors have to show up in complex conjugate pairs.
\item The $\mathfrak{e}(3)$ fixed point has 
only real eigenvalues.
There exist four relevant operators,
and three of them have almost $\Delta=2$ value.
It may be interesting to consider to define an interacting continuum field 
theory around it.
\end{itemize}

We make closer look at operators with $\Delta=2.000$ and those with complex $\Delta$. 
It is observed that when we estimate the fixed point values \eqref{eq:fpts_2-step}
without enough precision, two eigenvalues near 2.000 are not degenerate 
and eigenvectors belonging to them are in fact distinct. 
However, as the precision grows, both eigenvalues come close to 2.000 and 
eigenvectors also tend to coincide, and eventually we have single eigenvector 
with doubly degenerate eigenvalue 2.000 within much precision. Physically 
important question here is whether the eigenspace with the scaling dimension 2.000 
has dimension one or two, because it tells us the number of independence relevant operators 
with the scaling dimension of mass. The above observation seems to suggest that 
the eigenspace has dimension one and hence we have a single mass operators. 
At first sight, it seems strange that we have two types of the mass operators 
$\trN{\phi^2}$ and $\trN{\phi\phi^A}$  in the original action \eqref{action_AP}. 
In fact, it should be noticed that given sequence of matrices for which 
two eigenvalues tend to be degenerate and their eigenvectors tend to coincide, 
the dimension of the eigenspace in the limit is not always one.\footnote[2]
{For example, consider the sequence of matrices 
\begin{align}
\begin{pmatrix}
1 & -\epsilon^2 \\
\epsilon & 1 
\end{pmatrix},
\end{align}
with $\epsilon=1/m$ ($m \in\bm N$). Then it is easy to see for finite $m$ 
we have two distinct  eigenvalues each of which has a one-dimensional eigenspace, 
and that as $m\rightarrow\infty$, the eigenvalues and eigenvectors tend to agree. 
Nevertheless in the $m\rightarrow\infty$ limit 
we have a two-dimensional eigenspace of the eigenvalue 1.} 
Hence we have not had a definite answer to this question yet for lack of the exact value 
of the fixed points, but we note that the two types of the mass term  in the original action 
\eqref{action_AP} can be written in terms of the modes $\phi_{lm}$ as 
\begin{align}
&\frac{1}{N}\trN{\frac{\rho_N^2m_N^2}{2}\phi^2
+\frac{\rho_N^2\tilde m_N^2}{2}\phi\phi^A}
=\rho_N^2
\sum_{l,m}
\Big(\frac{m_N^2}{2}\phi_{lm}\phi^*_{lm}
+(-1)^l\frac{\tilde m_N^2}{2}\phi_{lm}\phi^*_{lm}\Big) \nn \\
&=\rho_N^2\sum_{l:\,\text{even},m}
\frac{m_N^2+\tilde m_N^2}{2}\phi_{lm}\phi^*_{lm}
+\rho_N^2\sum_{l:\,\text{odd},m}\frac{m_N^2-\tilde m_N^2}{2}\phi_{lm}\phi^*_{lm}.
\end{align}
Thus the two mass terms can be essentially regarded as the mass terms for even $l$ and 
odd $l$ modes. 
We do not expect that they are distinct operators with scaling dimension
 of mass.\footnote[3]{%
Since flipping the sign of $\tilde{m}_N^2$ is just exchanging the mass terms
for $l$ even and odd modes, this observation leads to a puzzle why there is no
fixed point with the sing flipped.
We do not have a definite answer now.} 
In fact, in the fixed points in terms of physical parameters \eqref{eq:fpts_2-step}
as well as the eigenvalues and eigenvectors  \eqref{evs}, 
we do not find any difference between even and odd $l$ and this fact ensures 
well-definedness of the large-$N$ limit. 
Hence it may be reasonable to conclude that we have only one operator with $\Delta=2$ 
in all cases $\mathfrak{e}(i)$ (or equivalently $\mathfrak{o}(i)$) for 
$i=0,\cdots, 3$. 

There exists a pair of complex eigenvalues in the fixed points $\mathfrak{e}(1)$, $\mathfrak{e}(2)$. As mentioned, the complex eigenvalue has to come in a pair;
$\lambda=1+w/N$ with $w=r e^{i\vartheta}$
and $w^*$ ($r\geq 0$ and $-\pi \leq \vartheta < \pi$).
By switching to a real eigenvector space, the linearized RGE matrix restricted 
to this eigenspace becomes
\begin{align}
  P\begin{pmatrix}
    1+\frac{w}{N} & 0 \\
    0 & 1+\frac{w^*}{N}
  \end{pmatrix}P^{-1}
=\bm 1_2+\frac{r}{N}
\begin{pmatrix}
   \cos\vartheta & - \sin\vartheta \\
   \sin \vartheta &  \cos \vartheta
\end{pmatrix}
\,.
\end{align}
for a unitary matrix $P$. 
Thus, in the space of real couplings, the corresponding flow of RGE transformation
(near the fixed point) is a spiral.
When it acts on a unit vector, the second part gives a vector of length $r/N \ll 1$.
If $\operatorname{Re}w>0$, namely $\cos\vartheta>0$, this tiny part is outward from the fixed points,
while when $\operatorname{Re}w<0$, i.e. $\cos\vartheta<0$, 
it points toward the fixed point.
Thus, the former is interpreted as a spiral source, and the latter is a spiral sink. 
When $\cos\varphi=0$ exactly, the flow of RGE transformation forms a limit cycle, 
and we need further subleading corrections in order to verify that this is the case.
In the present case the values of $\vartheta$ for each complex eigenvalues are
\begin{align}
  \mathfrak{e}(1): \quad
\vartheta=0.4925 = 0.1568\pi \,,
\qquad
  \mathfrak{e}(2): \quad
\vartheta=0.4760 = 0.1515\pi \,.
\end{align}
They are somehow close values, and correspond to spiral source behavior.

\subsubsection{Fixed points by original parameters and extra scaling factor}
\label{sec:fixed-points-orig}

So far, we have studied the properties of the fixed points
for the fuzzy sphere limit case ($\gamma=2$).
However, as we have stressed in the last part of Section
\ref{sec:fixed-points}, physical RGE flow has to be considered in
the physical parameter space spanned by $m_N^2, \tilde{m}_N^2, \kappa_N^{(a)}$.
When we look at the positions of the fixed points in this space
given by \eqref{def:rescaled_2step}, the finiteness of $x_*^{(i)}$ and $y_*^{(a)}$ 
in \eqref{xyfpeven} and \eqref{xyfpodd} implies that the fixed points are located 
in a finite region in the case of the fuzzy sphere limit,
where $\rho_N^2$ scales like $N^2$,
but they go to infinity in the NCFT limit, $N \rightarrow \infty$
with $\theta=2\rho_N^2/N$ fixed. The latter invalidates our perturbative calculation. 

In the case of well-known $d=4-\epsilon$ RG analysis in scalar field theory,
we need to include a nontrivial wave function renormalization factor
for a nontrivial fixed point to be realized.
If we canonically normalize the kinetic term, as we do here,
this extra factor gives rise a nontrivial $N$ dependence to the couplings.
In the previous study \cite{Kawamoto:2012ng},
we introduced an extra factor $c(N)=c N^\chi$ attached to the coupling constant
and tuned $\chi$ to realize the NCFT limit with a finite coupling constant.
Now, we try to introduce the same extra factor and reconsider the NCFT limit.
We replace the coupling constants as
\begin{align}
  \kappa_M^{(a)} \rightarrow c(M)  \kappa_M^{(a)} \,,
 \qquad
 (M=N-2\,,N-1\,,N)\,,
\label{eq:c_N_replace}
\end{align}
From \eqref{def:rescaled_2step}, one can easily see that $\chi$ should
be 1 for $\kappa_*^{(a)}$ to be finite in the NCFT limit.
Then it is obvious that we cannot keep $m_*^2$ and $\tilde{m}_*^2$ finite
by $c(N)$, but as long as all $\kappa^{(a)}_*$ remain finite and small, 
our perturbative approach is justified. 
From this observation, we also see that we can use the common $c(N)$ factor
for all the couplings.
In RGEs \eqref{eq:m_2-step}--\eqref{eq:kappa_a_2-step},
 we make the replacement \eqref{eq:c_N_replace},
and do the analysis again.
We now introduce a bit different rescaling variables,
\begin{align}
  m_*^2 =&
 \frac{\xi x_*^{(1)}}{\rho_N^2 P_N^*} \,,\qquad
  \tilde{m}_*^2 =
 \frac{\xi x_*^{(2)}}{\rho_N^2 P_N^*} \,,\qquad
\kappa_*^{(a)}= \frac{\xi }{c(N)}
\frac{b_N^2 b_{N-1}^2-1}{ b_N^2b_{N-1}^2 B_2(N
)\rho_N^2} y_*^{(a)}
\label{rescaled_2step_cN}
\,,
\end{align}
where
\begin{align}  
\xi =& \frac{b_N^2 b_{N-1}^2-\frac{c(N-2)}{c(N)}}{b_N^2 b_{N-1}^2-1}
 = \frac{\gamma+\chi}{\gamma} + \mathcal{O}(N^{-1})
 \,.
\end{align}
Note that in the NCFT limit of our interest, $\gamma=1$ and $\chi=1$,
and then $\xi= 2+\mathcal{O}(N^{-1})$, while in the fuzzy sphere limit, 
$\gamma=2$ and $\chi=0$, and so $\xi=1$, where \eqref{rescaled_2step_cN} 
is reduced to \eqref{def:rescaled_2step} (when $c=1$).
One can check that by use of these variables, the fixed point equations take the
same form as \eqref{fixed_point_eq_2step}.
Thus, we can use the same set of numerical solutions of ${\cal O}(1)$ given in 
\eqref{xyfpeven} and \eqref{xyfpodd}.

It should be noted that the propagator factor is modified too,
\begin{align}
P_N^* =& \frac{1-\xi \big(x_*^{(1)} + (-1)^{N-1} x_*^{(2)} \big) }{N(N-1)}
\,, \nn \\
\tilde{P}_{N-1}^* =&
\frac{1}{N(N-1)}\frac{1-\xi (x_*^{(1)} + (-1)^{N-1} x_*^{(2)})}
{1- 2\xi  (-1)^{N-1} x_*^{(2)}
-\frac{2}{N}\big(1-\xi(x_*^{(1)} + (-1)^{N-1} x_*^{(2)})\big) 
}
\,,
\label{def_P_N_xi}
\end{align}
and so are $B_1^*(N)$ and the other factors correspondingly. 
These equations with $\xi=1$ reproduce \eqref{eq:P*} and \eqref{eq:P*'}. 
This modification does affect the 
position of the fixed points in terms of the physical parameters.
By use of \eqref{rescaled_2step_cN}, we find
\begin{align}
&    \big(\theta m_*^2/N,\theta \tilde m_*^2/N,c\theta\kappa_*^{(0)}, c\theta\kappa_*^{(1)},
c\theta\kappa_*^{(2\alpha)},c\theta\kappa_*^{(2\beta)}\big)
\nn\\=&
\begin{cases}
  \big( -0.4006, 0.4006, 0.1367, -0.5466, 0.06885, 0.3411 \big)
& \text{
$\mathfrak{e}(1)$ solution} \,,\\
\big( -0.1982, 0.3643, 0.08557, -0.6291, 0.1711, 0.4925 \big)
& \text{
$\mathfrak{e}(2)$ solution} \,,\\
\big( -0.4861, -0.5156, 0.02998, 0.1272, 0.05996, 0.03748  \big)
& \text{
$\mathfrak{e}(3)$ solution} \,,\\
\big( -0.5010, 0.5010, 0.2137, -0.8548, 0.1077, 0.5334  \big)
& \text{
$\mathfrak{o}(1)$ solution} \,,\\
\big( -0.2423, 0.4454, 0.1279, -0.9404, 0.2558, 0.7363  \big)
& \text{
$\mathfrak{o}(2)$ solution} \,,\\
\big( -0.3865, -0.4100, 0.01895, 0.08040, 0.03790, 0.02369 \big)
& \text{
$\mathfrak{o}(3)$ solution} \,,
\end{cases}
\label{NCFTfps}
\end{align}
Therefore, the fixed point values do not match for even and odd $N$ in this case.
This may imply that this modification is not capable of defining theory around
fixed points in a well-defined manner, but we keep moving on to the linearized analysis.
At first sight it looks strange that the fixed point values of $m^2$ and $\tilde m^2$ are divergent 
in the large-$N$ limit as ${\cal O}(N)$. However, as discussed in \cite{Chu:2001xi}, 
this is the limit that leads to a massive scalar field theory with the well-known phase factor 
associated with each vertex on the Moyal plane. 

The linearized RGE \eqref{lin_RGE1} is also modified as
\begin{align}
    \delta \bm v_{N-2} =& \bigg( b_N^2 b_{N-1}^2 
    \begin{pmatrix}
      \mathbf{1}_2 & \\
              & \frac{c(N)}{c(N-2)} \mathbf{1}_4
    \end{pmatrix}
+ \bar{M} \bigg) \delta \bm v_N \,,
\end{align}
where $\bar{M}$ is modified from $\tilde{M}$ as
\begin{align}
  \bar{M} =&
  \begin{pmatrix}
    \mathbf{1}_2 & \\
           & \frac{1}{c(N-2)} \mathbf{1}_4
  \end{pmatrix}
  \begin{pmatrix}
    \xi \tilde{M}_{ij} & \tilde{M}_{ib} \\
    \xi^2 \tilde{M}_{aj} & \xi \tilde{M}_{ab}
  \end{pmatrix}
  \begin{pmatrix}
    \mathbf{1}_2 & \\
           & c(N) \mathbf{1}_4
  \end{pmatrix}
\,,
\label{def:M_bar}
\end{align}
where $i,j=1,2$ and $a,b=0,1,2\alpha,2\beta$, and $\tilde{M}_{ij}$ and the others are
the matrix elements given in \eqref{def:tilde_M}.
Note that the factors $B_i^*(N)$ and $\tilde{B}_i(N-1)$ are also modified
as in \eqref{def_P_N_xi}.
We again set the eigenvalue $\lambda=1+w/N$.
It is not difficult to see that, to the leading order in the $1/N$ expansion, 
we can drop the first and the third matrices in the definition of $\Bar{M}$ \eqref{def:M_bar}.
Thus, we can carry out numerical study as before.
It should be noted that we cannot forget these matrices when we fix the eigenvectors.
Actually, to obtain well-defined eigenvectors, we need to include $1/N$ corrections.
Thus, in the following we present the calculated values of the eigenvalues $w$,
for even and odd $N$ cases respectively.

For even $N$, 
\begin{align}
& \mathfrak{e}(1): \quad
w= 4.000  ,\,  2.044 \pm  0.5627 i ,\, 2.000 ,\, 0.9568 ,\, -5.513 ,\,   \nn\\
& \mathfrak{e}(2): \quad
w= 4.000 ,\, 2.618 ,\, 2.077 ,\,   1.235 ,\,  -0.3694 ,\,
-4.642 ,\, \nn\\
& \mathfrak{e}(3): \quad
w=  4.000 ,\,  3.470 ,\, 2.296 ,\, 2.000 ,\,  0.9338 ,\, 
-17.76 ,
\end{align}
and when $N$ is odd,
\begin{align}
& \mathfrak{o}(1): \quad
w=4.000 ,\,  2.000 ,\, 0.4404 \pm 1.683 i ,\, -2.865 ,\, -16.51 ,\, 
 \nn\\
& \mathfrak{o}(2): \quad
w=4.000 ,\,  1.988 ,\,0.9761 \pm 1.082 i ,\,  -3.180 ,\,  -10.88 ,\, 
\nn\\
& \mathfrak{o}(3): \quad
w=4.000 ,\,  3.769 ,\, 2.667 ,\, 2.339 ,\,  2.000 ,\,  -5.673 ,\,
\end{align}
and for the common Gaussian fixed point ($\mathfrak{e}(0)$ and $\mathfrak{o}(0)$),
the eigenvalues are $w=2$ (double)  and $w=4$ (quadruple).
Thus, we can observe that the scaling dimensions associated with fixed points
do not match for even and odd $N$.
They indeed share some properties; there are one $\Delta=4$ and one $\Delta\simeq 2$
operators for each case, for example, but
the scaling dimensions should be universal and are expected to coincide
for even and odd $N$.  
Together with the fact that the location of the fixed points does
not agree, we conclude that this modification utilizing $c(N)$ factor 
will not lead well-defined fixed points.

Although it might be possible to find a well-defined NCFT limit by
choosing $c(N)$ in a more elaborate way,
it is worth pointing out that at least perturbatively the antipode transformation is incompatible 
with the NCFT limit. Actually in this limit we have to restrict ourselves to the representation space of $SU(2)$ 
with $J_3=-(N-1)/2+{\cal O}(1)$ corresponding to a region near the south pole of the fuzzy sphere 
\cite{Chu:2001xi}. 
Hence it is evident that even if the original matrix $\phi$ is in this space, $\phi^A$  inevitably does not belong 
to it. In fact, we cannot figure out a possible NCFT limit of the most simple antipode interaction 
\eqref{eq:phiphiA}. Since in \eqref{NCFTfps} we found the fixed points in the NCFT limit with physical 
coupling constants as small as we can trust perturbation theory (at least by choosing $c$ appropriately), 
this is a problem which should be addressed even perturbatively. We can attribute the lack of 
consistent fixed points and scaling dimensions to this problem. 
In \cite{Chu:2001xi} it is confirmed that in the NCFT limit we have been discussing, 
a nonplanar diagram on the fuzzy sphere reproduces the well-known phase factor on the Moyal plane 
and as a consequence it has IR divergence via the UV/IR mixing \cite{Minwalla:1999px}. 
Since in the RG we look at the IR physics, it is very likely that non-existence of a well-defined fixed points 
in the NCFT limit reflects the UV/IR mixing. It is indeed true that the antipode interactions originate 
from the loop of the highest modes and we could not find any nontrivial fixed point including them 
in the NCFT limit in contrast to the fuzzy sphere limit. It would be interesting to examine more 
how the UV/IR mixing appears in our RG, in particular, in a nonperturbative manner.

\section{Conclusions and discussions}
\label{sec:discussions}

In this paper, we have formulated the large-$N$ renormalization group (RG)
for the rank $N$ matrix model which defines a scalar field theory on a fuzzy sphere. 
As a result of coarse-graining procedure in the large-$N$ RG, 
there (inevitably) appears an antipode field, which is defined as a scalar field twisted
by a sign factor fluctuating with respect to its angular momentum. 
The antipode field is characteristic of noncommutative nature of the geometry. 
For example, it would not emerge if we regularize the theory by use of simple cutoff. 
Thus according to the spirit of the RG, we start from 
the action with it describing new nonlocal interactions 
between fields and antipode ones.   

It has been discussed that the appearance of the antipode field spoils 
the validity of the RG structure \cite{Vaidya:2001bt}.
On the other hand, it has been also shown \cite{Chu:2001xi} 
that by integrating momenta of intermediate states over the whole range, 
the renormalized action becomes a smooth function of the external momenta.
Furthermore, such a smooth function gives rise to a characteristic
phase factor related to UV/IR mixing effects under a suitable limit
to noncommutative field theory.
These observations lead us to expectation that the RG analysis provides well-defined fixed point theory 
that would correspond to a field theory on a fuzzy sphere. 
Actually it turns out that the renormalization group equation (RGE) contains the oscillating phase factor 
$(-1)^{N-1}$ and that the fixed points given by these RGEs are somehow pathological.
It gives several continuum series of critical points, critical lines for even $N$, 
but it only provides isolated four points in the case of odd $N$.
The only common fixed point is Gaussian one, and properties
of Gaussian fixed point are also different with respect to $N$.
Thus, we may not expect that RGE is well defined in this case.
However, the arguments above suggest that 
this undesirable behavior would possibly be cured if we iterate
the RG transformation to include the effects of integrating out
lower momentum modes.
Hence we consider an RG transformation from the original size $N\times N$ 
to $(N-2)\times(N-2)$ one as a next step. 
This can be obtained by repetition of the RG transformation just given,
but to the lower order in perturbation theory it agrees with
the RGE defined by integrating out $l=2L$ and $2L-1$ modes. 
A fixed point of the RGE in this case describes the large-$N$ limit with keeping 
evenness or oddness of $N$. 
In fact, we find four fixed points for even and odd $N$ case respectively,
and also confirm that these four points are at the same locations.
We further carry out a linearized analysis around them, 
and also observe that they provide equivalent linearized theories 
with the same scaling dimensions and the set of scaling operators.
This is consistent with the claim in \cite{Chu:2001xi}, because 
we take account of the integration over lower modes by looking at the fixed points 
and then obtain the well-defined large-$N$ limit on the fuzzy sphere. 

More precisely, we consider two types of large-$N$ limits; the fuzzy sphere limit 
in which the fundamental scale of the fuzzy sphere is kept, 
and the NCFT limit that corresponds to zooming up a point on a fuzzy sphere to obtain
a noncommutative plane. 
In the case of the NCFT limit, we try to make the locations of the fixed points in a region 
where perturbation theory is valid by introducing an extra $N$ dependence to couplings. 
This however does not work well; the positions of fixed points are altered
differently for even and odd $N$ cases, and are no longer the same.
The linearized analysis also results in an inconsistent outcome for even and odd $N$.
Thus, this trial may not lead to a nice NCFT limit. 
This would reflect the IR singularity due the UV/IR mixing on the noncommutative plane, 
because the  antipode field originates from the loop of UV modes 
and resulting interactions between fields on the antipode points are IR phenomena. 
Thus we find sharp contrast with the fuzzy sphere limit and the NCFT limit,
and 
this observation could be regarded as a (nonperturbative) manifestation of the claim 
made in \cite{Chu:2001xi}, that the fuzzy sphere does not have the UV/IR mixing, 
but that the noncommutative anomaly there yields the UV/IR mixing in the NCFT limit. 

In the usual Wilsonian RG, the more we repeat the RG transformation,
the more kinds of interaction terms
we have.
In order to handle them, we usually argue that 
most of them are irrelevant and hence we could drop them. 
In the present case, we examine two-dimensional noncommutative field theories 
by using our large-$N$ RG. Thus we cannot invoke such argument. 
In fact, our RG also gives rise to several derivative corrections as shown explicitly 
in e.g. \eqref{eq:OoutOout}--\eqref{eq:OoutOout-OoutOout}. 
It is true that they are suppressed in $1/N$, are derivative terms written as the double commutator 
and hence can be neglected at least in low energy regime, but we should include them 
in the original action because they are actually generated via the RG. However, since we are 
in two dimensions, we have in principle infinitely many of them (because 
the scalar field has dimension zero) and it is impossible. Thus it is fair to say that 
we have found the fixed points and made analyses around them in the subspace 
of the coupling constants with such derivative corrections turned off. 

In our RG including antipode interactions, two well-defined fixed points have complex conjugate eigenvalues 
of the linearized RG transformation. This result is naturally interpreted as the fact 
that these fixed points are spiral sources in the two-dimensional subspace corresponding 
to operators associated with these eigenvalues. Such spiral behavior is quite rare 
in the ordinary Wilsonian RG in field theories. The reasons why we have it 
are explained as follows: in order to form a spiral flow, multiple operators need to be mixed in the RG. 
However, in the ordinary Wilsonian RG we consider operators with definite quantum numbers 
like the dimensions. It is then quite hard that operators with different quantum numbers are mixed. 
In contrast, in the present case we consider field theories in two dimensions 
in which a scalar field has the vanishing scaling dimension at least around the Gaussian fixed point. 
Moreover, the most essential reason would be that we have exact degeneracy of operators, 
i.e. a field and its antipode counterpart like $\phi$ and $\phi^A$. 
Their degeneracy is exact and is expected to hold even nonperturbatively 
as suggested by the property given in \eqref{eq:ref_AP_trace} in Appendix \ref{sec:matr-model-repr}. 
Thus they would be easily mixed and triggers the spiral behavior. Note that scaling dimensions 
around a fixed point are usually controlled by the conformal field theory, but in our case 
it is not available due to noncommutativity. 

The other nontrivial fixed point has real scaling dimensions
and eigenvectors.
On top of that, the fixed point values of the coupling constants
for the quartic terms are all positive,\footnote[4]{%
Note that the values of the masses
for the quadratic terms are negative, and
this is quite analogous to the situation in
usual $\phi^4$ theory in $D<4$.} in contrast to the other two fixed
points where there exist negative ones.
Thus it is possible that the theory around there can be well-defined,
and
the existence of such a fixed point may open a possibility that
we can define an interacting field theory in a noncommutative space
constructively.
Such field theory, if any, would provide us a hint what kind of
degrees of freedom emerges in the large-$N$ limit from the matrices.
This will be an interesting future direction.

It would also be interesting to examine how nonperturbative phenomena in the large-$N$ limit 
are captured in the large-$N$ RG. Among others, it is quite nice if 
supersymmetry breaking shown in \cite{Endres:2013sda} in the matrix model describing the lower dimensional 
superstring theory \cite{Kuroki:2013qpa} can be described via the large-$N$ RG.

\section*{Acknowledgments}
We thank C.~S.~Chu, H. Kawai, S. Iso, K. Ito, Y. Kimura, S. Moriyama,
H. Shimada, S. Shimasaki,
and F. Sugino 
for fruitful discussions and comments.
The work of S.~K. was supported in part by NSC103--2811--M--033--004
and NSC 103-2119-M-007-003.
S.~K. would like to thank NCTS, Physics Division where a part of the work
has been carried out. The work of T.~K. was supported in part by a Grant-in-Aid for Scientific Research (C), 25400274.

\appendix 


\section{Scalar field theory on a fuzzy sphere and useful formulas}
\label{sec:matr-model-repr}

We briefly introduce scalar field theory on a fuzzy sphere, which can be represented
as a hermitian matrix model.
We also present some interesting features of antipode projection, introduced in
Section \ref{sec:review_KKT}, and also provide some formulas which are useful
in our computation.
We make the introduction concise.
Readers may refer to \cite{Kawamoto:2012ng} for more detailed introduction.

\subsection{Scalar field theory on a  fuzzy sphere}
\label{sec:scalar_FT_Fuzzy}

We consider the following real scalar field theory on $S^2$ of radius $\rho$,
\begin{align}
S=\int\frac{\rho^2d\Omega}{4\pi}\left(-\frac{1}{2\rho^2}\left({\cal L}_i\phi(\theta,\varphi)\right)^2
+\frac{m^2}{2}\phi(\theta,\varphi)^2+\frac{g}{4}\phi(\theta,\varphi)^4\right) 
\,,
\label{eq:S2_action}
\end{align}
where a derivative operator ${\cal L}_i=-i\epsilon_{ijk}x_j\partial_k$ is related to
the Laplacian on a unit $S^2$ as
\begin{align}
  {\cal L}^2=-\Delta_{S^2}
= -\bigg(\frac{1}{\sin\theta}\frac{\partial}{\partial\theta}\sin\theta\frac{\partial}{\partial\theta}
+\frac{1}{\sin^2\theta}\frac{\partial^2}{\partial\varphi^2}
 \bigg) \,.
\label{def:S2_Laplacian}
\end{align}
The field $\phi(\theta, \varphi)$ can be expanded by use of the spherical harmonics
as
\begin{align}
\phi(\theta,\varphi)=\sum_{l=0}^{\infty}\sum_{m=-l}^l\plm Y_{lm}(\theta,\varphi) \,,
\label{fdecomp}
\end{align} 
and the action can be written in terms of the modes $\phi_{lm}$.
The reality condition implies that $\phi_{lm}^* = \phi_{l\,-m}$.
The spherical harmonics can be represented by use of symmetric traceless tensor
$c^{(lm)}_{i_1 \cdots i_l}$ as
\begin{align}
\ylm(\theta,\varphi)=\rho^{-l}\sum_{i_1 \cdots i_l}c^{(lm)}_{i_1 \cdots i_l}
x^{i_1}\cdots x^{i_l},
\label{yexp}
\end{align}
where $x^i$ ($i=1,\cdots , 3$) are the standard flat coordinate of $\bm R^3$.
$\ylm^*=(-1)^mY_{l\,-m}$ implies 
that ${c^{(lm)*}_{i_1\cdots i_l}}=(-1)^mc^{(l\,-m)}_{i_1\cdots i_l}$,
and the parity property $Y_{lm}(\pi-\theta, \varphi+\pi)= (-1)^l Y_{lm}(\theta,\varphi)$
is obvious from this expression since it corresponds to $x^{i_j} \rightarrow -x^{i_j}$
for $j=1,\cdots, l$.

Now, we introduce a fuzzy sphere.
Let $L_i$ ($i=1,2,3$) be the generators of spin $L=(N-1)/2$ representation of $SU(2)$.
We define $N \times N$ matrices $\hat{x}^i = \alpha L_i$, where $\alpha$ is a parameter
of length dimension.
To retain the relation $\sum_i (\hat{x}^i)^2 = \rho^2$, $\alpha$ is related to $\rho$ by
$\rho^2 = \alpha^2 (N^2-1)/4$. Notice that 
\begin{align}
[\hat x^i,\hat x^j]=i\alpha\epsilon_{ijk}\hat x^k, 
\label{eq:alpha}
\end{align}
which implies that $\alpha$ parametrizes noncommutativity on the fuzzy sphere. 
Using these $\hat{x}^i$, we can define $N \times N$ matrices $T_{lm}$, which we call the
fuzzy spherical harmonics, as
\begin{align}
\tlm=\rho^{-l}\sum_{i_1 \cdots i_l}c^{(lm)}_{i_1 \cdots i_l}
\hat x^{i_1}\cdots \hat x^{i_l} \,.
\label{texp}
\end{align}
Its hermitian conjugate is $T_{lm}^\dagger = (-1)^m T_{lm}$.
Let $\ket{s}$ be an $N$ dimensional representation space of $T_{lm}$ with $-L \leq s \leq L$.
The matrix element of $T_{lm}$ can be determined by Wigner-Eckart theorem, up to a normalization.
As in \cite{Kawamoto:2012ng}, we use the normalization of \cite{Iso:2001mg}
\begin{align}
(T_{lm})_{ss'}= 
\bra{s}T_{lm}\ket{s'}
=(-1)^{L-s}
\bep
L & l & L \\
-s & m & s'
\eep
\sqrt{(2l+1)N} \,,
\label{Tdef}
\end{align}
where the middle factor in the parenthesis is the Wigner's $3j$ symbol.
The orthogonality and the completeness thus follow,
\begin{align}
&\frac1N\trn{N}{\tlm\tnd{}^{\dagger}}=\delta_{ll'}\delta_{mm'}, 
\label{Tortho}\\
&
\frac1N \sum_{lm} 
(\tlm)_{s_1 s_2} (\tlm^\dagger)_{s_3 s_4}
=\delta_{s_1s_4}\delta_{s_2s_3} \,,
\label{Tcomp}
\end{align}
and $\tlm$ spans a complete basis for $N\times N$ matrices.
Finally, corresponding to the Laplacian operator on $S^2$ \eqref{def:S2_Laplacian},
we have
\begin{align}
  -\Delta \phi \equiv  [L_i,[L_i,\phi]] \,,
\quad
\text{where} \quad
-\Delta T_{lm} = l(l+1) \tlm \,.
\label{matrixLaplacian}
\end{align}
We thus define a mapping rule\footnote[3]{Similar mapping rule for noncommutative superspace 
is given in \cite{Kawai:2003yf}.} from a scalar field theory on $S^2$ to
$N\times N$ matrix counter part as
\begin{enumerate}
\item function $\rightarrow$ matrix:
\begin{align}
\phi(\theta,\varphi)=\sum_{l=0}^{2L}\sum_{m=-l}^l\plm Y_{lm}(\theta,\varphi)
\quad \rightarrow \quad
\phi=\sum_{l=0}^{2L}\sum_{m=-l}^l\plm \tlm.
\label{mapping}
\end{align}
\item integration $\rightarrow$ trace:
\begin{align}
\int\frac{d\Omega}{4\pi}\phi(\theta,\varphi)=\frac1N\tr_N\phi. 
\label{intandtr}
\end{align}
Notice that this holds as equality. 
\item Laplacian $\rightarrow$ double adjoint action:
\begin{align}
-\Delta_{\Omega}\phi(\theta,\varphi) \quad \rightarrow \quad \Lap{\phi}.
\label{Lapandadj}
\end{align}
\end{enumerate}
By following these rules, scalar field theory on a fuzzy sphere that corresponds to
\eqref{eq:S2_action} is defined as
\begin{align}
S=\frac{\rho^2}{N}\trn{N}{-\frac{1}{2\rho^2}[L_i,\phi]^2+\frac{m^2}{2}\phi^2+\frac{g}{4}\phi^4}\,.
\end{align}
By putting the subscript $N$ to $\rho$ and $m^2$ for convenience
and replacing
$g$ with $\kappa_N^{(0)}$, we have our starting action \eqref{action}. 
Note that according to the mapping rule \eqref{mapping} and by using the fusion 
of $T_{lm}$ given in \eqref{eq:fusion2} later, the matrix product $\phi_1\phi_2$,
with $\phi_i=\sum_{l_im_i}(\phi_i)_{l_im_i}T_{l_im_i}$
($i=1,2$),
corresponds to a noncommutative product of fields on $S^2$ given as 
\begin{align}
\phi_1(\theta,\varphi) \ast \phi_2(\theta,\varphi)
=\sum_{l_i\,m_i\,(i=1,\cdots , 3)}
\left((\phi_1)_{l_1m_1}{F_{l_1m_1\,l_2m_2}}^{l_1,m_3}(\phi_2)_{l_2m_2}\right)
Y_{l_3m_3}(\theta,\varphi),
\label{star}
\end{align}
where 
${F_{l_1m_1\,l_2m_2}}^{l_1,m_3}$ is shown in \eqref{eq:fusion2}.
From this expression,
we recognize that this star product is indeed noncommutative because the fusion 
${F_{l_1m_1\,l_2m_2}}^{l_1,m_3}$ is not symmetric under interchange 
between $(l_1,m_1)$ and $(l_2,m_2)$.

\subsection{Useful formulas of the fuzzy spherical harmonics}
\label{app:tlm}
In this appendix useful formulas of the $N\times N$ matrix $\tlm$ 
defined in \eqref{Tdef} are presented.

First, the orthogonality \eqref{Tortho} and the completeness \eqref{Tcomp} lead to
\begin{align}
&\frac1N\sum_{lm}\trn{N}{\mathcal{O}_1\tlm}\trn{N}{\mathcal{O}_2\tlm^{\dagger}}
=\trn{N}{\mathcal{O}_1\mathcal{O}_2}, 
\label{merging}\\
&\frac1N\sum_{lm}\trn{N}{\mathcal{O}_1\tlm \mathcal{O}_2\tlm^{\dagger}}=\tr_N\mathcal{O}_1
\, \tr_N \mathcal{O}_2,
\label{splitting}
\end{align}
for arbitrary $N\times N$ matrices $\mathcal{O}_1$, $\mathcal{O}_2$. 
These properties are used to combine double traces into single traces.

By use of the formulas for the $3j$, $6j$, and $9j$ symbols \cite{Var},
the traces
of $\tlm$ 
can be evaluated as
\begin{align}
&\trn{N}{\tn{1}\tn{2}\tn{3}}
=N^{\frac32}\prod_{i=1}^3(2l_i+1)^{\frac12}(-1)^{2L+\sum_{i=1}^3l_i}
\bep
l_1 & l_2 & l_3 \\
m_1 & m_2 & m_3
\eep
\beB
l_1 & l_2 & l_3 \\
L & L & L
\eeB,
\label{3T} \nn \\
&\trn{N}{\tn{1}\tn{2}\tn{3}\tn{4}} \\
&=N^2\prod_{i=1}^4(2l_i+1)^{\frac{1}{2}}\sum_{lm}(-1)^{-m}(2l+1) 
\bep
l_1 & l_4 & l \\
m_1 & m_4 & m
\eep
\bep
l & l_3 & l_2 \\
-m & m_3 & m_2
\eep
\beB
l_1 & l_4 & l \\
L & L & L
\eeB
\beB
l & l_3 & l_2 \\
L & L & L
\eeB 
\label{4T1}
\\
&=N^2\prod_{i=1}^4(2l_i+1)^{\frac{1}{2}}(-1)^{l_2+l_3}\sum_{lm}(-1)^{l-m}(2l+1) 
\bep
l_1 & l_3 & l \\
m_1 & m_3 & m
\eep
\bep
l & l_2 & l_4 \\
-m & m_2 & m_4 
\eep
\beB
l_1 & l_3 & l \\
L & L & l_2 \\
L & L & l_4
\eeB. 
\label{4T2}
\end{align}
The curly brackets with six and nine entries are the $6j$ 
and $9j$ symbols,respectively.
The relevant formulas are also summarized in \cite{Kawamoto:2012ng}.

{}From \eqref{Tortho} and \eqref{3T}, we can derive the
following useful fusion formula,
\begin{align}
  T_{l_1m_1} T_{l_2m_2} =&\sum_{l_3\,m_3}{F_{l_1 m_1\,l_2 m_2}}^{l_3 m_3}T_{l_3m_3}, 
\nn \\
{F_{l_1 m_1\,l_2 m_2}}^{l_3 m_3}=&N^{\frac12}\prod_{i=1}^3(2l_i+1)^{\frac12}
(-1)^{2L+\sum_{i=1}^3l_i+m_3}
\bep
l_1 & l_2 & l_3 \\
m_1 & m_2 & -m_3
\eep
\beB
l_1 & l_2 & l_3 \\
L & L & L
\eeB
\,.
\label{eq:fusion2}
\end{align}

By using this twice, one can easily derive a 
``similarity transformation'' property by $T_{2L-n\; m}$,
\begin{align}
&  \sum_{m=-2L+n}^{2L-n} (-1)^m T_{2L-n \; m} T_{l_1 m_1} T_{2L-n \; -m}
=
N(2N-1-2n)
\begin{Bmatrix}
   L & L & l_1 \\
  L  & L  & 2L-n
\end{Bmatrix}
(-1)^{n+l_1}
T_{l_1 m_1} \,.
\label{eq:similarity}
\end{align}
If the matrix sandwiched by $T_{2L-n\; m}$ and $T_{2L-n\; m}^\dagger$
is an identity (namely $l_1=0$), one finds
\begin{align}
  \sum_{m=-2L+n}^{2L-n} (-1)^m T_{2L-n \; m} \mathbf{1}_N T_{2L-n \; -m}
=&
(2N-1-2n) \mathbf{1}_N \,.
\end{align}

\subsection{Properties of the antipode projection}
\label{sec:prop-antip-proj}

In this appendix, we show that the antipode projection
enjoys the following property:
\begin{align}
\label{adj_prop_AP}
  (\phi_1 \phi_2 \cdots \phi_m)^A = \phi_m^A \cdots \phi_2^A \phi_1^A \,.
\end{align}
Namely, the antipode of a string of operators is the string of the opposite ordering
with antipode operators.
\begin{flushleft}
  \underline{Proof:}
\end{flushleft}
For two operators $(\phi \psi)^A$, by using the fusion formula \eqref{eq:fusion2},
\begin{align}
&  (\phi \psi)^A 
\nn\\=& 
N^{\frac12}\prod_{i=1}^3(2l_i+1)^{\frac12}
(-1)^{2L+\sum_{i=1}^3l_i+m_3}
\bep
l_1 & l_2 & l_3 \\
m_1 & m_2 & -m_3
\eep
\beB
l_1 & l_2 & l_3 \\
L & L & L
\eeB
\phi_{l_1 m_1} \psi_{l_2 m_2}
T_{l_3m_3}^A
\nn\\=&
\psi^A \phi^A \,.
\end{align}
Thus,
\begin{align}
  \left(\phi_1 \phi_2 \cdots \phi_m \right)^A
= \left(\phi_2 \cdots \phi_m \right)^A \phi_1^A
=\cdots =
\phi_m^A \cdots \phi_2^A \phi_1^A \,,
\end{align}
which concludes the proof.
\\

From  $\trn{N}{T_{lm}} = N \delta_{l0} \delta_{m0}$,
it immediately follows that $  \tr (\phi^A) = \tr ( \phi)$.
This leads to the following reflection property of the antipode projection
inside a trace,
\begin{align}
\label{eq:ref_AP_trace}
  \tr_N \big( \phi_1 \cdots \phi_n \big) =&
\tr_N \big( \phi_n^A \cdots \phi_1^A \big) \,.
\end{align}
Namely, we can reverse the ordering of the fields inside a trace by
putting the antipode projection to all of them.

\subsection{Useful formulas of $3nj$-symbols}
\label{app:3nj}

In this appendix, we summarize useful formulas we use our perturbative
calculations.
The details require more formulas than collected here, but
the readers may refer to the previous paper \cite{Kawamoto:2012ng},
or the textbook \cite{Var}. 
Many of the asymptotic relations for the $6j$ symbols here are derived 
by applying the Stirling's formula to Racah's exact expression of the $6j$ symbols.

\subsubsection{Asymptotic formulas and shift relations of $6j$ symbols}
\label{sec:asymptotic-formulas}

\paragraph{asymptotic formulas}

Racah's asymptotic formula: for $a,b,c,\gg f$,
\begin{align}
  \begin{Bmatrix}
    a & b & c \\
    b & a & f 
  \end{Bmatrix}
=&
  \begin{Bmatrix}
    a & a & f \\
    b & b & c 
  \end{Bmatrix}
\simeq
\frac{(-1)^{a+b+c+f}}{\sqrt{(2a+1)(2b+1)}}
P_f \left(\cos\theta \right) \,,\\
\cos\theta =& \frac{a(a+1)+b(b+1)-c(c+1)}{2\sqrt{a(a+1)b(b+1)}} \,.
\end{align}
Thus, for $n,l \ll L=(N-1)/2$,
\begin{align}
  \begin{Bmatrix}
    L & L & l \\
    L & L & 2L-n
  \end{Bmatrix}
=&
\frac{(-1)^n}{2L+1}
\bigg[
1 - \frac{l(l+1)(2n+1)}{2L}
+\mathcal{O}(L^{-2})
\bigg]
\nn\\=&
\frac{(-1)^n}{N}
\left[
1-\frac{l(l+1)}{N} (2n+1)
+{\cal O}(N^{-2})
\right] \,.
\label{eq:6j_asymp1}
\end{align}

For $n,m \ll L$, by use of the Racah formula and Stirling's formula,
\begin{align}
&  \begin{Bmatrix}
    L & L & 2L-n \\
    L & L & 2L-m
  \end{Bmatrix}
\nn\\=&
2^{-4L-2} \sqrt{\frac{2\pi}{L}}
\cdot 4^{n-m} L^{n-2m}
 \sum_{t=0}^m
\frac{(-1)^{t-m} (16L^3)^t n!m!}{[(t+n-m)!]^2(t!)^2(m-t)!}
\left( 1 + {\cal O}\big(L^{-1} \big) \right)
\label{eq:6j_asymp2}
\,,
\end{align}
where we have assumed $n \geq m$ without loss of generality due to
the symmetry of the $6j$ symbol.
In the similar way, for $m \ll L$, one can find
\begin{align}
\beB
2L-m & 2L & 2L \\
L & L & L
\eeB
=& \frac{(-1)^{2L-m} 3^{\frac{3}{4}}(2\pi)^\frac14}{8 \sqrt{m!}}L^{\frac{m}{2}-\frac34}\left(\frac34\right)^{3L-\frac{m}{2}}
\big( 1+ \mathcal{O}(L^{-1}) \big)
\,.
\label{eq:6j_asymp3}
\end{align}

\paragraph{shift of the argument}

If $R\gg 1$ and $a$, $b$, $c$ are arbitrary, 
\begin{align}
(-1)^{2R}
\beB
a & b & c \\
d+R & e+R & f+R
\eeB
\simeq
\frac{(-1)^{c+d+e}}{\sqrt{2R(2c+1)}}C_{a(f-e)\,b(d-f)}^{c(d-e)}, 
\label{p306(4)}
\end{align}
where $C_{a\alpha\,b\beta}^{c\gamma}$ is the Clebsch-Gordan coefficient\footnote[4]
{The formula presented in \cite{Var} needs a phase factor given here.}
whose relation to the $3j$ symbol is
\begin{align}
  C^{c\gamma}_{a\alpha b\beta} =&
(-1)^{a-b+\gamma} \sqrt{2c+1}
\begin{pmatrix}
 a & b & c \\
 \alpha & \beta & -\gamma
\end{pmatrix}
\,.
\end{align}
When $a+b+c$ is an even number, the Clebsch-Gordan coefficient satisfies the following
shift property,
\begin{align}
  a+b+c=\text{even}: \qquad
C^{c0}_{a1b-1} = C^{c0}_{a-1b1} =
\frac{c(c+1) - a(a+1) - b(b+1)}{2\sqrt{a(a+1)b(b+1)}}
C^{c0}_{a0b0} \,.
\label{eq:CG_shift}
\end{align}
By using \eqref{p306(4)},
one can derive the following relation
\begin{align}
  \begin{Bmatrix}
    l & l_1 & l_2 \\
    L   & L  & L \pm 1
  \end{Bmatrix}
\simeq
\frac{C^{l_2 0}_{l\pm 1 \; l_1 \mp 1}}{C^{l_2 0}_{l 0 l_1 0}}
  \begin{Bmatrix}
    l_1 & l_2 & l_3 \\
    L   & L  & L
  \end{Bmatrix}
\,.
\label{eq:6j_shift}
\end{align}
Since this is based on the asymptotic relation \eqref{p306(4)},
it holds only to the leading order in the large-$L$ limit.
When $l+l_1+l_2$ is even, by \eqref{eq:CG_shift}, one finds
\begin{align}
  \begin{Bmatrix}
    l & l_1 & l_2 \\
    L   & L  & L \pm 1
  \end{Bmatrix}
\simeq
\frac{l_2(l_2+1) - l(l+1) - l_1(l_1+1)}{2\sqrt{l(l+1)l_1(l_1+1)}}
  \begin{Bmatrix}
    l & l_1 & l_2 \\
    L   & L  & L
  \end{Bmatrix}
\label{eq:6j-shift}
\,.
\end{align}

\subsubsection{Asymptotic formula for $9j$ symbols}
\label{sec:asympt-form-9j}

We consider an asymptotic expansion formula for a $9j$ symbol,
\begin{align}
  \begin{Bmatrix}
  l_1 & l_2 & l \\
  L  & L  & 2L-m \\
  L & L & 2L-n
\end{Bmatrix}
\,,  
\end{align}
where $l,l_1,l_2,m,n \ll L$.
The symmetry of the $9j$ symbol suggests that this is invariant under the simultaneous
exchange of $l_1 \leftrightarrow l_2$ and $n \leftrightarrow m$.
The basic strategy to derive a formula is the same as \cite{Kawamoto:2012ng},
namely by use of the decomposition into $6j$ symbols,
\begin{align}
&
  \begin{Bmatrix}
  l_1 & l_2 & l \\
  L  & L  & 2L-m \\
  L & L & 2L-n
\end{Bmatrix}
=
\sum_X (-1)^{2X} (2X+1)
\begin{Bmatrix}
  l & l_1 & l_2 \\
  L & L & X
\end{Bmatrix}
\begin{Bmatrix}
  L & X & l \\
  2L-m & 2L-n & L
\end{Bmatrix}
\begin{Bmatrix}
  L & L & l_1 \\
  L & X & 2L-m
\end{Bmatrix}
\,,
\end{align}
where the triangular relations of $6j$ symbols imposes the conditions,
\begin{align}
&  L-\text{min}(l,l_1,m) \leq X \leq L+\text{min}(l,l_1)
\,,
\quad
|m-n| \leq l \leq 2L \ll 4L-m-n \,,
\nn\\ &
\text{max}(  m-l,0) \leq n \leq m+l \,,
\quad
\text{max}(  n-l,0) \leq m \leq n+l \,,
\end{align}
and the usual ones for $l,l_1$, and $l_2$.
The last two $6j$ symbols are evaluated by use of the exact expression \`a la
the Racah and Stirling's formula, as presented in Appendix B in \cite{Kawamoto:2012ng}.
In the previous study, we need only $n=m=0$ case.
In this case, we need to take care of the range of the summation in Racah's formula
and the calculation is much more complicated.
We thus work out only for the case $0 \leq n,m \leq 1$,
and present the result,
\begin{align}
&
  \begin{Bmatrix}
  l_1 & l_2 & l \\
  L  & L  & 2L-n \\
  L & L & 2L-n
\end{Bmatrix}
\nn\\=&
\frac{(-1)^{N-1+l} \sqrt{2}}{2N}
\begin{Bmatrix}
  l & l_1 & l_2 \\
  L & L & L
\end{Bmatrix}
\bigg[
1-\frac{2n+1}{4} \frac{2l_1(l_1+1) + 2l_2(l_2+1) - l(l+1) -1}{N}
+\mathcal{O}(N^{-2})
\bigg]
\,,
\label{eq:9j_asymp1}
\\
&
\begin{Bmatrix}
  l_1 & l_2 & l \\
  L  & L  & 2L \\
  L & L & 2L-1
\end{Bmatrix}
=
\frac{(-1)^{2L+l+1} \sqrt{2}N }{16 L^{5/2}}
\bigg[
\sqrt{l(l+1)}
\begin{Bmatrix}
  l & l_1 & l_2 \\
  L & L & L
\end{Bmatrix}
+2\sqrt{l_1(l_1+1)}
\begin{Bmatrix}
  l & l_1 & l_2 \\
  L & L & L+1
\end{Bmatrix}
\bigg]
\,,
\label{eq:9j_asymp2}
\end{align}
where $n=0,1$.
The case with $m=1$ and $n=0$ can be obtained by exchanging $l_1$ and $l_2$ in
the second formula.
In the second formula, the shift relation \eqref{eq:6j-shift} implies that
these two $6j$ symbols are of the same order.
Thus, the second $9j$ with $n=1$ and $m=0$ itself is subleading compared to the first one,
namely $n=m=0,1$ cases.

\section{Calculations of the expectation values}
\label{app:calc_VEVs}

In this appendix we evaluate generic forms of the expectation values that are necessary
to derive renormalization group equations.
We first provide general expressions for expectation values
by integrating out $2L, 2L-1 , \cdots, 2L-\hat{n}+1$ out modes
as in \eqref{def:mom_shell_n-step}.
It is convenient to define a part of the out modes
with the angular momentum $l=2L-n$ as
\begin{align}
  \phout_{n} =& \sum_{m} \phout_{2L- n \; m} T_{2L-n \; m} \,,
\end{align}
where $n \ll N$.
Then the out mode field \eqref{def:in-out-fields} can be written as
$\phout= \sum_{n=0}^{\hat{n}-1} \phout_n$.
Note that from the action \eqref{action_AP} the propagator is still of diagonal form,
\begin{align}
&  \vev{\phi^\text{out}_{2L-n_1 m_1} \phi^\text{out}_{2L-n_2 m_2} }_0=
\delta_{n_1 n_2} \delta_{m_1+m_2} (-1)^{m_1} P_{N-n_1} \,,
\nn\\ &
P_{N-n}= \frac{1}{(N-n)(N-n-1)[1+(-1)^{N-n-1}\zeta_{N-n}]
+\rho_{N-n}^2[m_{N-n}^2 + (-1)^{N-n-1}\tilde{m}_{N-n}^2]} \,.
\end{align}

We first consider
\begin{align}
\vev{\tr_N \left({\cal O}_1 \; \phout \; {\cal O}_2 \; \phout \right)}_0
=&
\sum_{n,m=0}^{\hat{n}-1}  \vev{\tr_N \left({\cal O}_1 \; \phout_n \; {\cal O}_2 \; \phout_m \right)}_0
\,,
\end{align}
where 
$\mathcal{O}_i$ ($i=1,2$) are generic polynomials of $\phin$ and $\phin^A$.
It is sufficient to calculate the following piece,
\begin{align}
&  \vev{\tr_N \left({\cal O}_1 \; \phout_n \; {\cal O}_2 \; \phout_{n'} \right)}_0
\nn\\=&
\delta_{nn'}  P_{N-n} 
\sum_{m} (-1)^m \tr_N \left( {\cal O}_1 \; T_{2L-n\; m} {\cal O}_2 \; T_{2L-n \; -m}  \right)
\nn\\=&
\delta_{nn'}  N(2N-1-2n) \; P_{N-n} 
\sum_{l,m} 
\begin{Bmatrix}
  L & L & l \\
  L & L & 2L-n
\end{Bmatrix}
(-1)^{n+l}
({\cal O}_2)_{lm} \tr_N \big({\cal O}_1  T_{lm} \big)
\nn\\=&
\delta_{nn'}
 (2N-1-2n) \; P_{N-n} 
\tr_N \bigg[
{\cal O}_1 {\cal O}_2^A
-\frac{2n+1}{N} {\cal O}_1 (-\Delta) {\cal O}_2^A
+{\cal O}(N^{-2})
\bigg] \,,
\label{eq:1st_1}
\end{align}
where we have used \eqref{eq:similarity} and \eqref{eq:6j_asymp1}.
In the final form,
$\mathcal{O}_2^A$ can be evaluated by use of a property of the antipode
projection \eqref{adj_prop_AP}.
$-\Delta \mathcal{O}$ stands for $[L_i, [L_i, \mathcal{O}]]$ 
introduced in \eqref{matrixLaplacian}. 

Next we consider
\begin{align}
&  \vev{\tr_N \big({\cal O}_1 \phout {\cal O}_2 \phout \big)
\tr_N \big(\phout \big)^4}_c
\nn\\=&
\sum_{n_3, \cdots, n_8=0}^{\hat{n}-1}
\vev{\tr_N \big({\cal O}_1 \phout_{n_3} {\cal O}_2 \phout_{n_4} \big)
\tr_N \big(\phout_{n_5} \phout_{n_6} \phout_{n_7} \phout_{n_8} \big)}_c
\,.
\label{eq:mass_1}
\end{align}
We need to evaluate connected graphs, which are divided into
\begin{align}
&
\vev{\tr_N \big({\cal O}_1 \phout_{n_3} {\cal O}_2 \phout_{n_4} \big)
\tr_N \big(\phout_{n_5} \phout_{n_6} \phout_{n_7} \phout_{n_8} \big)}_c
\nn\\=& 
4 \delta_{n_3 n_6} \delta_{n_4 n_5} \delta_{n_7 n_8}
\vev{\tr_N \big({\cal O}_1 \phout_{n_3} {\cal O}_2 \phout_{n_4} \big)
\tr_N \big(\phout_{n_4} \phout_{n_3} \phout_{n_5} \phout_{n_5} \big)}_c
\label{eq:mass_1_1}
\\
+&
4\delta_{n_3 n_5} \delta_{n_4 n_6} \delta_{n_7 n_8}
\vev{\tr_N \big({\cal O}_1 \phout_{n_3} {\cal O}_2 \phout_{n_4} \big)
\tr_N \big(\phout_{n_3} \phout_{n_4} \phout_{n_5} \phout_{n_5} \big)}_c
\label{eq:mass_1_2}
\\
+&
4\delta_{n_3 n_7} \delta_{n_4 n_5} \delta_{n_6 n_8}
 \vev{\tr_N \big({\cal O}_1 \phout_{n_3} {\cal O}_2 \phout_{n_4} \big)
\tr_N \big(\phout_{n_4} \phout_{n_5} \phout_{n_3} \phout_{n_5} \big)}_c
\label{eq:mass_1_3}
\,,
\end{align}
where the out modes with the same index $n_i$ are to be contracted.
We evaluate them separately.
It is easy to see from the following calculation that \eqref{eq:mass_1_1} and \eqref{eq:mass_1_2} give the same answer, and we evaluate it as
\begin{align}
  &
\vev{\tr_N \big({\cal O}_1 \phout_{n_3} {\cal O}_2 \phout_{n_4} \big)
\tr_N \big(\phout_{n_4} \phout_{n_3} \phout_{n_5} \phout_{n_5} \big)}_c
\nn\\=&
P_{N-n_3}P_{N-n_4}P_{N-n_5}
\sum_{m_3,m_4,m_5} (-1)^{m_3+m_4+m_5}
\nn\\& \hskip3em \times
\tr_N \big( {\cal O}_1 T_{2L-n_3\,m_3} {\cal O}_2 T_{2L-n_4\,m_4} \big)
\tr_N \big( T_{2L-n_4\, -m_4} T_{2L-n_3\,  -m_3} T_{2L-n_5\,m_5} T_{2L-n_5 \,-m_5}  \big)
\nn\\=&
\delta_{n_3n_4}  \big( P_{N-n_3} \big)^2 P_{N-n_5}
N^2 (2N-1-2n_5)(2N-1-2n_3)
\nn\\&\hskip4em \times
\sum_{l,m} 
\begin{Bmatrix}
  L & L & l \\
  L & L & 2L-n_3
\end{Bmatrix}
(-1)^{n+l}
({\cal O}_2)_{lm} \tr_N \big({\cal O}_1  T_{lm} \big)
\nn\\=&
\delta_{n_3n_4} 
N \big( P_{N-n_3} \big)^2 
(2N-1-2n_3)
\cdot
P_{N-n_5}
(2N-1-2n_5)
\nn\\&\hskip5em \times
\tr_N \bigg[
{\cal O}_1 {\cal O}_2^A
-\frac{2n_3+1}{N} {\cal O}_1 (-\Delta) {\cal O}_2^A
+{\cal O}(N^{-2})
\bigg] \,,
\end{align}
where we have again used \eqref{eq:similarity} and \eqref{eq:6j_asymp1}.
On the other hand, the rest term \eqref{eq:mass_1_3} is
\begin{align}
  &
 \vev{\tr_N \big({\cal O}_1 \phout_{n_3} {\cal O}_2 \phout_{n_4} \big)
\tr_N \big(\phout_{n_4} \phout_{n_5} \phout_{n_3} \phout_{n_5} \big)}_c
\nn\\=&
\delta_{n_3n_4}
 \big( P_{N-n_3} \big)^2 P_{N-n_5}
N^2 (2N-1-2n_5) 
\sum_{m_3} (-1)^{m_3}
\nn\\& \hskip2em \times
\begin{Bmatrix}
  L & L & 2L-n_3 \\
  L & L & 2L-n_5
\end{Bmatrix}
(-1)^{2L-n_3+n_5}
\tr_N \big( {\cal O}_1 T_{2L-n_3 m_3} {\cal O}_2 T_{2L-n_3\;  -m_3} \big)
\,.
\end{align}
According to the asymptotic formula \eqref{eq:6j_asymp2}, this is exponentially
small for large-$N$.

Finally, we consider
\begin{align}
&
\vev{ \tr_N \left({\cal O}_1 \phout {\cal O}_2 \phout \right)
\tr_N \left({\cal O}_3 \phout {\cal O}_4 \phout  \right)}_c
\nn\\=&
\sum_{n_5,\cdots,n_8=0}^{\hat{n}-1}
\vev{ \tr_N \left({\cal O}_1 \phout_{n_5} {\cal O}_2 \phout_{n_6} \right)
\tr_N \left({\cal O}_3 \phout_{n_7} {\cal O}_4 \phout_{n_8}  \right)}_c
\,.
\end{align}
Again, the connected pieces are
\begin{align}
&
\vev{ \tr_N \left({\cal O}_1 \phout_{n_5} {\cal O}_2 \phout_{n_6} \right)
\tr_N \left({\cal O}_3 \phout_{n_7} {\cal O}_4 \phout_{n_8}  \right)}_c
\nn\\=&
\delta_{n_5 n_8}\delta_{n_6 n_7}
\vev{ \tr_N \left({\cal O}_1 \phout_{n_5} {\cal O}_2 \phout_{n_6} \right)
\tr_N \left({\cal O}_3 \phout_{n_6} {\cal O}_4 \phout_{n_5}  \right)}_c  
\\
&+
\delta_{n_5 n_7}\delta_{n_6 n_8}
\vev{ \tr_N \left({\cal O}_1 \phout_{n_5} {\cal O}_2 \phout_{n_6} \right)
\tr_N \left({\cal O}_3 \phout_{n_5} {\cal O}_4 \phout_{n_6}  \right)}_c  
\,.
\end{align}
Due to the cyclic symmetry of the trace, the former is equal to
the latter with ${\cal O}_3 \leftrightarrow {\cal O}_4$.
We evaluate the latter as
\begin{align}
&
\vev{ \tr_N \left({\cal O}_1 \phout_{n_5} {\cal O}_2 \phout_{n_6} \right)
\tr_N \left({\cal O}_3 \phout_{n_5} {\cal O}_4 \phout_{n_6}  \right)}_c  
\nn\\=&  
P_{N-n_5} P_{N-n_6}
\sum_{m_5,m_6} (-1)^{m_5+m_6}
\sum_{l_i,m_i} \prod_{i=1}^4 ({\cal O}_i)_{l_im_i}
\nn\\& \hskip2em \times
\tr_N \left(T_{l_1 m_1} T_{2L-n_5 m_5} T_{l_2 m_2} T_{2L-n_6 m_6} \right)
\tr_N \left(T_{l_3 m_3} T_{2L-n_5\; -m_5} T_{l_4 m_4} T_{2L-n_6\; -m_6}\right)
\nn\\=&  
N^4 P_{N-n_5} P_{N-n_6} (4L-2n_5+1)(4L-2n_6+1)
\sum_{m_5,m_6} (-1)^{m_5+m_6}
\sum_{l_i,m_i} \prod_{i=1}^4 ({\cal O}_i)_{l_im_i} \sqrt{2l_i+1}
\nn\\& \hskip2em \times
(-1)^{2L-n_5+l_2 + 2L-n_6+l_4}
\sum_{\tilde{l},\tilde{m},\tilde{l}',\tilde{m}'}
(-1)^{\tilde{l}-\tilde{m}+\tilde{l}'-\tilde{m}'} (2\tilde{l}+1)(2\tilde{l}'+1)
\nn\\& \hskip2em \times
\begin{pmatrix}
  l_1 & l_2 & \tilde{l} \\
  m_1 & m_2 & \tilde{m}
\end{pmatrix}
\begin{pmatrix}
  \tilde{l} & 2L-n_5 & 2L-n_6 \\
  -\tilde{m} & m_5 & m_6
\end{pmatrix}
\begin{pmatrix}
  l_3 & l_4 & \tilde{l}' \\
  m_3 & m_4 & \tilde{m}'
\end{pmatrix}
\begin{pmatrix}
  \tilde{l}' & 2L-n_5 & 2L-n_6 \\
  -\tilde{m}' & -m_5 & -m_6
\end{pmatrix}
\nn\\& \hskip2em \times
\begin{Bmatrix}
  l_1 & l_2 & \tilde{l} \\
  L  & L  & 2L-n_5 \\
  L & L & 2L-n_6
\end{Bmatrix}
\begin{Bmatrix}
  l_3 & l_4 & \tilde{l}' \\
  L  & L  & 2L-n_5 \\
  L & L & 2L-n_6
\end{Bmatrix}
\nn\\=&  
N^4 P_{N-n_5} P_{N-n_6} (2N-2n_5-1)(2N-2n_6-1)
\sum_{l_i,m_i} \prod_{i=1}^4 ({\cal O}_i)_{l_im_i} \sqrt{2l_i+1}
\nn\\& \hskip2em \times
(-1)^{l_2 +l_4}
\sum_{\tilde{l},\tilde{m}}
(-1)^{\tilde{l}-\tilde{m}} (2\tilde{l}+1)
\begin{pmatrix}
  l_1 & l_2 & \tilde{l} \\
  m_1 & m_2 & \tilde{m}
\end{pmatrix}
\begin{pmatrix}
  l_3 & l_4 & \tilde{l} \\
  m_3 & m_4 & -\tilde{m}
\end{pmatrix}
\nn\\& \hskip2em \times
\begin{Bmatrix}
  l_1 & l_2 & \tilde{l} \\
  L  & L  & 2L-n_5 \\
  L & L & 2L-n_6
\end{Bmatrix}
\begin{Bmatrix}
  l_3 & l_4 & \tilde{l} \\
  L  & L  & 2L-n_5 \\
  L & L & 2L-n_6
\end{Bmatrix}
\,.
\end{align}
As noted in Appendix \ref{sec:asympt-form-9j},
we do not have a general asymptotic formula for
this $9j$ symbol at hand.
Only available ones are those with $0 \leq n_5, n_6 \leq 1$;
namely we can only evaluate the $\hat{n}=2$ case.
As shown by \eqref{eq:9j_asymp1} and \eqref{eq:9j_asymp2}, the leading order contributions
are from $n_5=n_6=0$ and $1$. The result is
\begin{align}
&
\vev{ \tr_N \left({\cal O}_1 \phout_{n} {\cal O}_2 \phout_{n} \right)
\tr_N \left({\cal O}_3 \phout_{n} {\cal O}_4 \phout_{n}  \right)}_c
\nn\\=& 
 \frac{NB_2(N-n)}{2} 
\tr_N \bigg[
\mathcal{O}_1^A \mathcal{O}_2 \mathcal{O}_4^A \mathcal{O}_3
-\frac{2n+1}{2N}
\bigg(
-\sum_i \Delta^{(i)} \big(\mathcal{O}_1^A \mathcal{O}_2 \mathcal{O}_4^A \mathcal{O}_3 \big)
+ \mathcal{O}_1^A \mathcal{O}_2 \Delta \big( \mathcal{O}_4^A \mathcal{O}_3 \big)
\bigg)
\nn\\& \hskip8em
+(\mathcal{O}_3 \leftrightarrow \mathcal{O}_4)
+\mathcal{O}(N^{-2})
\bigg]
\,,
\end{align}
where $n=0$ and $1$, and $B_2(N-n)=2(2N-2n-1)\big( P_{N-n} \big)^2$ is used.
Note that $\Delta^{(i)}$ acts only on $\mathcal{O}_i$.
$n_5 \neq n_6$ case is $1/N$ suppressed compared to these contributions.

Finally, we cite the calculation of $\vev{\mathcal{V}_3^{(0)}\mathcal{V}_3^{(0)}}_0$
from (C.16) of \cite{Kawamoto:2012ng},
\begin{align}
  &
\vev{\tr_N (\phin \phout^3) \tr_N(\phin \phout^3)}_c
\nn\\=&
P_N^3(2N-1)^3N^4\sum_{(l_1,m_1),(l'_1,m'_1)\in\Li}\pin{1}\pind{1}
\sqrt{(2l_1+1)(2l'_1+1)} \nn \\
&\times\sum_{l,l'}(2l+1)(2l'+1)
\beB
l_1 & 2L & l \\
L & L & L
\eeB
\beB
l'_1 & 2L & l' \\
L & L & L 
\eeB
\beB
l & 2L & 2L \\
L & L & L
\eeB
\beB
l' & 2L & 2L \\
L & L & L 
\eeB \nn \\
&\times
\sum_{m_1,\cdots , m_3}\sum_{m,m'}(-1)^{m_1+m_2+m_3-m-m'}
\bep
l_1 & 2L & l \\
m_1 & m_4 & m 
\eep
\bep
l & 2L & 2L \\
-m & m_3 & m_2 
\eep \nn \\
&\times 
\left[
\bep
l'_1 & 2L & l' \\
m'_1 & -m_2 & m'
\eep
\bep
l' & 2L & 2L \\
-m' & -m_3 & -m_4
\eep
+(\text{permutations of}~m_2,\cdots , m_4)
\right].
\end{align}
The triangular conditions for the first two $6j$ symbols
impose $l=2L-m$ and $l'=2L-n$ with $m \leq l_1 \ll L$
and $n \leq l_1' \ll L$.
According to the asymptotic formula of $6j$ symbols \eqref{eq:6j_asymp3},
this contribution is indeed exponentially suppressed for large $L$,
as argued in \eqref{eq:V3V3-result}.
We  can therefore drop this term from our perturbative calculation.


\end{document}